\documentclass[a4paper,12pt]{article}
\usepackage{amsmath}
\usepackage{amssymb}
\usepackage{amsfonts}
\usepackage{amssymb,amsmath}
\usepackage{cancel}
\usepackage[dvips]{lscape,graphicx,epsfig}
\usepackage{fleqn}
\usepackage[footnotesize]{caption}
\usepackage{mathrsfs}

\def\beq{\begin{equation}}
\def\eeq{\end{equation}}
\def\bea{\begin{eqnarray}}
\def\eea{\end{eqnarray}}

\def\<{\left\langle}
\def\>{\right\rangle}

\renewcommand{\baselinestretch}{1.30}

\newcommand{\bc}{\begin{center}}
\newcommand{\ec}{\end{center}}
\newcommand{\bd}{\begin{displaymath}}
\newcommand{\ed}{\end{displaymath}}
\newcommand{\be}{\begin{equation}}
\newcommand{\ee}{\end{equation}}
\newcommand{\ba}{\begin{array}}
\newcommand{\ea}{\end{array}}
\newcommand{\bt}{\begin{tabular}}
\newcommand{\et}{\end{tabular}}

\begin{document}

\bibliographystyle{OurBibTeX}
\begin{titlepage}
\vspace*{-15mm}
\vspace*{5mm}

\begin{center}
{
\sffamily
\Large Leptogenesis in the Exceptional Supersymmetric Standard Model:
flavour dependent lepton asymmetries
}
\\[8mm]
S.F.~King$^{a}$,
R.~Luo$^{b}$,
D.J.~Miller$^{b}$ and
R.~Nevzorov$^{b}$\footnote{On leave of absence from the Theory Department, ITEP, Moscow, Russia}
\\[3mm]
{\small\it
$^a$ School of Physics and Astronomy, University of Southampton,\\
Southampton, SO17 1BJ, U.K.\\[2mm]
$^b$ Department of Physics and Astronomy, University of Glasgow,\\
Glasgow G12 8QQ, U.K.
}\\[1mm]
\end{center}
\vspace*{0.75cm}

\begin{abstract}
\noindent We calculate flavour dependent lepton asymmetries within
the $E_6$ inspired Supersymmetric Standard Model ($\rm E_6SSM$),
which has an extra $U(1)_N$ symmetry. In this  model, the
right-handed neutrino doesn't participate in gauge  interactions,
allowing it to be used for both the see--saw mechanism and
leptogenesis. Extra Higgs, leptons and leptoquarks
predicted by the E$_6$SSM contribute
to the ordinary lepton CP asymmetries induced by the decays of the
lightest right--handed neutrino (and sneutrino) and give rise to a set
of extra decay asymmetries. We find that the CP asymmetries can be
relatively large, even when the lightest right--handed neutrino is
as light as $10^6\,\mbox{GeV}$.
\end{abstract}

\end{titlepage}

\newpage

\setcounter{footnote}{0}

\section{Introduction}

The baryon asymmetry in the universe
$\eta_{B}=(n_{B}-n_{\bar{B}})/n_{\gamma}=(6.21\pm 0.16)\times
10^{-10}$ \cite{Komatsu:2008hk} is one of the motivations to
explore physics beyond the elementary particle Standard Model
(SM), if the Big Bang nucleosynthesis theory is the right
description of the evolution of the universe. Among new physics
mechanisms for baryogenesis: are GUT baryogenesis
\cite{gut-baryogen}, electroweak baryogenesis
\cite{ew-baryogen}--\cite{Huber:2000mg}, baryogenesis via
leptogenesis \cite{Fukugita:1986hr}--\cite{review-leptogen}, the
Affleck-Dine mechanism \cite{Affleck-Dine} and so on\,.
Leptogenesis is a particularly interesting mechanism for the
generation of baryon asymmetry. It is almost unavoidable in seesaw
models \cite{see-saw} because all three Sakharov conditions
\cite{Sakharov:1967dj} can be naturally fulfilled in this
scenario. In particular, the seesaw mechanism requires that lepton
number is violated while complex neutrino Yukawa couplings can
provide a source for CP violation. In the simplest realisation of
the seesaw mechanism, i.e. in the so--called type I seesaw models,
the lepton asymmetry is induced by the heavy right-handed neutrino
decays. This lepton asymmetry is partially converted into baryon
asymmetry via sphaleron processes \cite{Kuzmin:1985mm}.

A lot of work has been done in the scenario of the leptogenesis from right-handed neutrino decay.
Initially lepton CP asymmetries which stem from decays of the lightest right-handed neutrino were calculated
within the SM \cite{CPasym-SM} and its minimal supersymmetric (SUSY) extension (MSSM) \cite{CPasym-SUSY}
assuming the type I seesaw mechanism of neutrino mass generation. In the early studies of leptogenesis (see for example
\cite{Buchmuller:2004nz}) flavour effects were ignored. The importance of flavour effects was emphasised in
\cite{CPleptogen-flavour}, \cite{Antusch:2006cw}. The process of lepton asymmetry generation in the seesaw models
with triplet scalar field and/or triplet fermion field was analysed in \cite{triplet-seesaw-leptogen1}--\cite{triplet-seesaw-leptogen2}.
In the case of Dirac neutrinos leptogenesis was studied in \cite{dirac-leptogen}.

A potential drawback of supersymmetric thermal leptogenesis is the lower bound on the mass of the lightest right-handed
neutrino $M_1$. Indeed, it was shown that the appropriate amount of baryon asymmetry in the SM and MSSM can be induced only
if $M_1\gtrsim 10^9\,\mbox{GeV}$ \cite{lower-bound}. In the framework of supergravity the lower bound on $M_1$
leads to the gravitino problem \cite{gravitino-problem} as follows. After inflation the universe thermalizes with a reheat
temperature $T_R$. If $T_R> M_1$, right-handed neutrinos are produced by thermal scattering in the reheating epoch. This means that
thermal leptogenesis could take place in the MSSM and other SUSY models if $T_R\gtrsim 10^9\,\mbox{GeV}$.
At the same time such a high reheating temperature results in an overproduction of gravitinos. Because gravitinos have a long
lifetime they may decay during or after Big Bang Nucleosynthesis (BBN) destroying the agreement between the predicted and observed light
element abundances. Hence the relic abundance of gravitinos should be constrained from above to preserve the success of BBN.
It was argued that the gravitino density becomes low enough when $T_R\lesssim 10^{6-7}\,\mbox{GeV}$ \cite{Kohri:2005wn}.
For such a low reheating temperature and relatively small mass of the lightest right-handed neutrino $M_1\sim TeV$
thermal leptogenesis in SUSY models can still be effective if a set of soft supersymmetry breaking terms provides an
additional source of both lepton number violation and CP violation (soft leptogenesis) \cite{soft-leptogenesis}, \cite{triplet-seesaw-leptogen2}.
$\mbox{TeV}$ scale thermal leptogenesis is also possible if the spectrum of heavy Majorana right-handed neutrinos is quasi--degenerate
(so--called resonant leptogenesis \cite{resonant-leptogen}) or the particle content of the considered models involves extra particles
beyond the SM and/or MSSM \cite{new-particles}. Alternatively, right-handed neutrinos can be produced non--thermally even
at very low reheating temperatures, for instance in inflaton decay \cite{inflaton-leptogen}, or in preheating \cite{Giudice:1999fb}.
The gravitino problem can be evaded automatically if the gravitino is the lightest supersymmetric particle in the considered SUSY model
\cite{gravitino-LSP} or if it is rather heavy so that gravitinos decay before BBN \cite{Ibe:2004tg}.

In this paper, we study the generation of lepton asymmetry in the framework of the Exceptional Supersymmetric
Standard Model (E$_6$SSM) \cite{King:2005jy}-\cite{King:2005my}. This $E_6$ inspired SUSY model is based on
the low--energy standard model gauge group together with an extra $U(1)_{N}$ gauge symmetry under which right-handed
neutrinos have zero charge. In the E$_6$SSM the $\mu$ problem is solved in a similar way to the NMSSM, but without
the accompanying problems of singlet tadpoles or domain walls. Because right--handed neutrinos do not participate
in the gauge interactions in the considered model they may be superheavy, shedding light on the origin of the mass
hierarchy in the lepton sector and providing a mechanism for the generation of lepton and baryon asymmetry of the Universe.

Our analysis presented here goes well beyond what has appeared so far in the literature \cite{Hambye:2000bn}. In this article
we calculate lepton CP asymmetries that stem from the decays of the lightest right-handed neutrino taking into account flavour effects.
It means that, in contrast with \cite{Hambye:2000bn}, we treat the decay asymmetries associated with different lepton flavours in
the final state separately. We also define and compute flavour CP asymmetries originating from the decays of the lightest right--handed
neutrino into the exotic leptoquarks (and squarks) that carry lepton and baryon number simultaneouly where our results differ from those in
\cite{Hambye:2000bn}. Finally we perform a comprehensive numerical analysis of the impact of the new particles and interactions
appearing in the E$_6$SSM on the CP asymmetries within see--saw models with sequential dominance of right--handed
neutrinos \cite{SD}--\cite{King:2002qh}. Our results show that it may be possible to achieve successful thermal leptogenesis
even with reheat temperature as low as $10^6\,\mbox{GeV}$.

The paper is organised as follows. In the next section we briefly review the E$_6$SSM. In section 3 we calculate lepton
CP asymmetries within this model. The results of our numerical analysis are discussed in section 4. Section 5 is reserved
for our conclusions and outlook.

\section{Exceptional SUSY model}

\noindent
The E$_6$SSM is based on the $SU(3)_C\times SU(2)_W\times U(1)_Y \times U(1)_N$ gauge group
which is a subgroup of $E_6$. An additional low energy $U(1)_N$, that is not present either
in the SM or in the MSSM, is a linear superposition of $U(1)_{\chi}$ and $U(1)_{\psi}$, i.e.
\be
U(1)_N=\dfrac{1}{4} U(1)_{\chi}+\dfrac{\sqrt{15}}{4} U(1)_{\psi}\,,
\label{lg1}
\ee
where $U(1)_{\psi}$ and $U(1)_{\chi}$ symmetries are defined by:
$$
E_6\to SO(10)\times U(1)_{\psi}\,,\qquad SO(10)\to SU(5)\times U(1)_{\chi}\,.
$$
To ensure anomaly cancellation the particle content of the E$_6$SSM is extended to
include three complete fundamental $27$ representations of $E_6$ at low energies.
These multiplets decompose under the $SU(5)\times U(1)_{N}$ subgroup of $E_6$ as follows:
\begin{equation}
\begin{array}{c}
27_i\to \left(10,\,\displaystyle\frac{1}{\sqrt{40}}\right)_i+
\left(5^{*},\,\displaystyle\frac{2}{\sqrt{40}}\right)_i
+\left(5^{*},\,-\displaystyle\frac{3}{\sqrt{40}}\right)_i +
\left(5,-\displaystyle\frac{2}{\sqrt{40}}\right)_i+\\[3mm]
+\left(1,\displaystyle\frac{5}{\sqrt{40}}\right)_i+\left(1,0\right)_i.
\end{array}
\label{lg2}
\end{equation}
The first and second quantities in the brackets are the $SU(5)$
representation and extra $U(1)_{N}$ charge while $i$ is a family index
that runs from 1 to 3. An ordinary SM family which contains the doublets of left-handed quarks
$Q_i$ and leptons $L_i$, right-handed up- and down-quarks ($u^c_i$ and $d^c_i$)
as well as right-handed charged leptons, is assigned to $\left(10,\dfrac{1}{\sqrt{40}}\right)_i
+\left(5^{*},\,\dfrac{2}{\sqrt{40}}\right)_i$. Right-handed neutrinos $N^c_i$ should be
associated with the last term in Eq.~(\ref{lg2}), $\left(1,0\right)_i$.
The next-to-last term in Eq.~(\ref{lg2}), $\left(1,\dfrac{5}{\sqrt{40}}\right)_i$, represents
SM-type singlet fields $S_i$ which carry non-zero $U(1)_{N}$ charges and therefore survive
down to the EW scale.  The pair of $SU(2)_W$--doublets ($H_{1i}$ and $H_{2i}$)
that are contained in $\left(5^{*},\,-\dfrac{3}{\sqrt{40}}\right)_i$ and
$\left(5,-\dfrac{2}{\sqrt{40}}\right)_i$ have the quantum numbers of
Higgs doublets. So they form either Higgs or inert Higgs $SU(2)_W$ multiplets
\footnote{We use the terminology "inert Higgs" to denote Higgs like doublets that
do not develop vacuum expectation values (VEVs).}.
Other components of these $SU(5)$ multiplets form colour triplets of exotic quarks
$\overline{D}_i$ and $D_i$ with electric charges $-1/3$ and $+1/3$ respectively.
These exotic quark states carry a $B-L$ charge $\left(\pm\dfrac{2}{3}\right)$ twice
larger than that of ordinary ones. Therefore in phenomenologically viable $E_6$ inspired
models they can be either diquarks or leptoquarks.

In addition to the complete $27_i$ multiplets the low energy particle spectrum of the E$_6$SSM is
supplemented by $SU(2)_W$ doublet $H'$ and anti-doublet $\overline{H}'$ states from extra $27'$
and $\overline{27'}$ to preserve gauge coupling unification. These components of the $E_6$ fundamental
representation originate from $\left(5^{*},\,\dfrac{2}{\sqrt{40}}\right)$ of $27'$ and
$\left(5,\,-\dfrac{2}{\sqrt{40}}\right)$ of $\overline{27'}$ by construction. The splitting of $27'$
and $\overline{27'}$ multiplets can be naturally achieved, for example, in the framework of orbifold
GUTs \cite{121}. Thus, in addition to a $Z'$ corresponding to the $U(1)_N$ symmetry, the E$_6$SSM involves
extra matter beyond the MSSM with the quantum numbers of three $5+5^{*}$ representations of $SU(5)$ plus
three $SU(5)$ singlets with $U(1)_N$ charges. The presence of a $Z'$ boson and exotic quarks predicted
by the E$_6$SSM provides spectacular new physics signals at the LHC which were discussed in
\cite{King:2005jy}-\cite{King:2005my}, \cite{Accomando:2006ga}.

As any other supersymmetric model, the E$_6$SSM suffers from problems related with
rapid proton decay. In other words gauge symmetry does not forbid lepton and
baryon number violating operators. Moreover exotic particles in the $E_6$ inspired SUSY
models give rise to new Yukawa interactions that induce unacceptably large non-diagonal
flavour transitions in general. To suppress flavour changing processes in the E$_6$SSM
an approximate $Z^{H}_2$ symmetry is imposed. All superfields except one pair of
$H_{1,i}$ and $H_{2,i}$ (say $H_d\equiv H_{1,3}$ and $H_u\equiv H_{2,3}$) and one SM-type singlet
field ($S\equiv S_3$) are odd under this symmetry. The $Z^{H}_2$ symmetry reduces the structure
of the Yukawa interactions to:
\be
\ba{c}
W_{\rm E_6SSM}\simeq  \lambda S (H_u H_d)+ \lambda_{\alpha\beta} S (H_{1\alpha} H_{2\beta})+
\kappa_{ij} S (D_i\overline{D}_j)+ f_{\alpha\beta}(H_d H_{2\alpha})S_{\beta}\nonumber\\[2mm]
+\tilde{f}_{\alpha\beta}(H_{1\alpha}H_u)S_{\beta}+h^U_{ij}(H_{u} Q_i)u^c_{j} + h^D_{ij}(H_{d} Q_i)d^c_j
+ h^E_{ij}(H_{d} L_i)e^c_{j}+ h_{ij}^N(H_{u} L_i)N_j^c\nonumber\\[2mm]
+\dfrac{1}{2}M_{ij}N^c_iN^c_j+\mu'(L_4 \overline{L}_4)+h^{E}_{4j}(H_d L_4)e^c_j+h_{4j}^N (H_{u} L_4)N_j^c\,.
\ea
\label{lg3}
\ee
where $L_4\equiv H'$, $\overline{L}_4\equiv \overline{H}'$, $\alpha,\beta=1,2$ and $i,j=1,2,3$\,.
One can notice that the survival components from the $27'$ and $\overline{27'}$ manifest themselves in the
Yukawa interactions (\ref{lg3}) as fields with lepton number $L=\pm 1$. Consequently, $L_4$ couples to other
fields as a fourth family lepton doublet and our notations reflect this. The $SU(2)_W$ doublets $H_u$ and $H_d$,
that are even under $Z^{H}_2$ symmetry, play the role of Higgs fields generating the masses of quarks and leptons
after electroweak symmetry breaking (EWSB). The extra $U(1)_{N}$ gauge symmetry forbids an elementary $\mu$ term
in the superpotential of E$_6$SSM but allows the interaction of Higgs doublets with the SM--type singlet field $S$.
The vacuum expectation value (VEV) of the field $S$ breaks the extra $U(1)_N$ symmetry thereby providing
an effective $\mu$ term as well as the necessary exotic fermion masses.

The superpotential of the E$_6$SSM includes two types of bilinear terms. One of them,
$\mu' L_4\overline{L}_4$, is solely responsible for the masses of the charged and neutral
components of $L_4$ and $\overline{L}_4$. The corresponding mass term is not suppressed by
$E_6$ and is not involved in the process of EWSB. Therefore the parameter $\mu'$ remains arbitrary.
Recent analysis revealed that gauge coupling unification in the E$_6$SSM is consistent with
$\mu'$ around $100\,\mbox{TeV}$ \cite{King:2007uj}. Another type of bilinear terms
$\dfrac{1}{2}M_{ij} N_i^c N_j^c$, determines the spectrum of the right--handed neutrinos.
These mass terms are forbidden by $E_6$ and can be generated only after its breakdown \cite{Howl:2008xz}.
Suppose $N^c_H$ and~ $\overline{N}_H^c$ are components of some extra $27_H$ and
$\overline{27}_H$ representations which develop VEVs along the $D$--flat direction
$\langle N_H^c \rangle = \langle \overline{N}_H^c \rangle \simeq\Lambda$. Then the right--handed neutrino mass terms can be
induced through the non--renormalisable interactions of $27_i$ and $\overline{27}_H$ of the form
$\dfrac{\eta_{ij}}{M_{Pl}}(\overline{27}_H\, 27_i)(\overline{27}_H\, 27_j)$.
As a result right--handed neutrinos gain masses $M_i$ of the order of
$\dfrac{\Lambda^2}{M_{Pl}}<<M_X$. At the same time we assume that $M_i$
are much larger than $\mu'$ so that the right--handed neutrinos can decay
either to a Higgs particle and a fermion component of $L_4$ or to a higgsino and a scalar
component of $L_4$.

Although $Z^{H}_2$ eliminates any problem related with non-diagonal flavour transitions it also forbids all
Yukawa interactions that would allow the exotic quarks to decay. Since models with stable charged exotic
particles are ruled out by various experiments \cite{Hemmick:1989ns} the $Z^{H}_2$ symmetry can only be
an approximate one. But the breakdown of $Z^{H}_2$ should not give rise to operators leading to rapid proton decay.
There are two ways to overcome this problem. The resulting Lagrangian has to be invariant either with respect to
an exact $Z_2^L$ symmetry, under which all superfields except lepton ones are even (Model I), or with respect to an exact
$Z_2^B$ discrete symmetry, under which exotic quark and lepton superfields are odd whereas the others
remain even (Model II). If the Lagrangian is invariant under the $Z_2^B$ symmetry transformations then exotic quarks are
leptoquarks. If $Z_2^L$ is imposed then the baryon number conservation requires the exotic quarks to be diquarks.
The breakdown of $Z^{H}_2$ symmetry also leads to the new interactions of the right--handed neutrinos with exotic
particles. The corresponding terms in the superpotential of the E$_6$SSM are given by
\begin{equation}
\Delta W=\xi_{\alpha ij}(H_{2\alpha} L_i)N_j^c+\xi_{\alpha4 j}(H_{2\alpha} L_4)N_j^c+
g^{N}_{kij}D_{k}d^{c}_{i}N^{c}_{j}\,.
\label{lg5}
\end{equation}
The Yukawa couplings $g^N_{ijk}$ vanish if exotic quarks are diquarks and may have non---zero values if
exotic quarks are leptoquarks. Because $Z^{H}_2$ symmetry violating interactions may give an appreciable
contribution to the amplitude of $K^0-\overline{K}^0$ oscillations and give rise to new muon decay channels
like $\mu\to e^{-}e^{+}e^{-}$ the Yukawa couplings of the related terms are required to be small ($\lesssim 10^{-3}-10^{-4}$).
This suggests that $\xi_{\alpha ij}$ should also be similarly small, but does not provide any constraint on the
couplings $\xi_{\alpha 4j}$.

Combining the appropriate terms in Eqs.~(\ref{lg3}) and (\ref{lg5}) one can write the part of the superpotential
describing the interactions of the right--handed neutrinos with other bosons and fermions in the following compact form:
\begin{equation}
W_N= h^N_{kxj}(H^u_{k} L_x)N_j^c + g^{N}_{kij}D_{k}d^{c}_{i}N^{c}_{j}\,,
\label{lg4}
\end{equation}
where $H^u_3\equiv H_u$ is the usual Higgs doublet,~ $H^u_{\alpha}\equiv H_{2\alpha}$
are the two extra inert Higgs doublets, $L_4$ is the extra lepton doublet, while
$D_k$ are the exotic quarks.
Note that the indices run over the following ranges ~ $x=1,2,3,4$~ while~ $k,i,j=1,2,3$.
In the Model I $g^{N}_{kij}=0$ so the only extra particles present
are inert Higgs and the fourth lepton doublet,
whereas in the Model II all terms which appear on the right--hand side of
Eq.~(\ref{lg4}) including extra leptoquarks can be present.

\section{Decay asymmetries in the ${\rm\bf E_6SSM}$}

\subsection{CP asymmetries for the Model I}
In this subsection we discuss Model I corresponding to the
case of additional inert Higgs and a fourth lepton family only.
The case of additional leptoquarks is considered in the next
subsection.

In models with heavy right--handed neutrinos lepton asymmetry can be dynamically generated and
then get converted into a baryon asymmetry due to $(B+L)$--violating sphaleron interactions.
The generation of lepton asymmetry occurs via the out--of equilibrium decay of the lightest
right--handed neutrino $N_1$. The process of the lepton asymmetry generation is controlled by the
flavour CP (decay) asymmetries $\varepsilon_{1,\,\ell_k}$ that appear on the right--hand side of
Boltzmann equations. In the SM there are three decay asymmetries associated with three lepton flavours
$e,\,\mu$ and $\tau$. They are given by
\begin{equation}
\varepsilon_{1,\,\ell_k}=\dfrac{\Gamma_{N_1 \ell_{k}}-\Gamma_{N_1 \bar{\ell}_{k}}}
{\sum_{m} \left(\Gamma_{N_1 \ell_{m}}+\Gamma_{N_1 \bar{\ell}_{m}}\right)}\,.
\label{lg6}
\end{equation}
where $\Gamma_{N_1 \ell_{k}}$ and $\Gamma_{N_1 \bar{\ell}_{k}}$ are
partial decay widths of $N_1\to L_k+H_u$ and $N_1\to \overline{L}_k+H^{*}_u$ with $k,m=1,2,3$.
At the tree level CP asymmetries (\ref{lg6}) vanish because $\Gamma_{N_1 \ell_{k}}=\Gamma_{N_1 \bar{\ell}_{k}}$.
If CP invariance is broken in the lepton sector the non--zero contributions to the CP asymmetries arise from
the interference between the tree--level amplitudes of the lightest right--handed neutrino decays and
one--loop corrections to them.

Supersymmetry gives rise to new channels of right--handed neutrino decay into sleptons $\widetilde{L}_k$ and Higgsino
$\widetilde{H}_u$ that also contribute to the generation of total lepton asymmetry. The corresponding flavour
CP asymmetries are defined as
\begin{equation}
\varepsilon_{1,\,\widetilde{\ell}_k}=\dfrac{\Gamma_{N_1 \widetilde{\ell}_{k}}-\Gamma_{N_1 \widetilde{\ell}^{*}_{k}}}
{\sum_{m} \left(\Gamma_{N_1 \widetilde{\ell}_{m}}+\Gamma_{N_1 \widetilde{\ell}^{*}_{m}}\right)}\,.
\label{lg12}
\end{equation}
In addition supersymmetry predicts the existence of a scalar partner of the right--handed neutrino $\widetilde{N}_1$
(right--handed sneutrino). The decays of the right--handed sneutrino into lepton and Higgsino and
into slepton and Higgs provide another important origin of lepton asymmetry.
The right--handed sneutrino CP asymmetries can be determined similarly to the neutrino ones
\begin{equation}
\varepsilon_{\widetilde{1},\,\ell_k}=\dfrac{\Gamma_{\widetilde{N}_1^{*} \ell_{k}}-\Gamma_{\widetilde{N}_1 \bar{\ell}_{k}}}
{\sum_{m} \left(\Gamma_{\widetilde{N}_1^{*} \ell_{m}}+\Gamma_{\widetilde{N}_1 \bar{\ell}_{m}}\right)}\,,\qquad
\varepsilon_{\widetilde{1},\,\widetilde{\ell}_k}=\dfrac{\Gamma_{\widetilde{N}_1 \widetilde{\ell}_{k}}-\Gamma_{\widetilde{N}_1^{*}
\widetilde{\ell}^{*}_{k}}}
{\sum_{m} \left(\Gamma_{\widetilde{N}_1 \widetilde{\ell}_{m}}+\Gamma_{\widetilde{N}_1^{*} \widetilde{\ell}^{*}_{m}}\right)}\,.
\label{lg13}
\end{equation}
The direct computation of decay asymmetries in SUSY models reveals that
\begin{equation}
\varepsilon_{1,\,\ell_k}=\varepsilon_{1,\,\widetilde{\ell}_k}=\varepsilon_{\widetilde{1},\,\ell_k}=
\varepsilon_{\widetilde{1},\,\widetilde{\ell}_k}\,.
\label{lg15}
\end{equation}

In the Exceptional SUSY model the relation between different types of decay asymmetries (\ref{lg15}) remains intact.
But extra particles predicted by the E$_6$SSM result in the new channels of the decays of right--handed neutrino and
its superpartner. In the E$_6$SSM Model I only inert Higgs superfields and $L_4$ are allowed to have non--zero
Yukawa couplings to the right--handed neutrino superfields (see Eq.~(\ref{lg4})). Since the extra
inert Higgs and the fourth family of (vector-like) leptons are
expected to be significantly lighter than $N_1$ these Yukawa interactions
induce new decay modes of the lightest right--handed neutrino and sneutrino. A complete set of possible decay
channels of the lightest right--handed neutrino and sneutrino includes
\begin{equation}
N_1\to L_x + H^u_k,\qquad N_1\to \widetilde{L}_x+\widetilde{H}^u_k,\qquad
\widetilde{N}_1\to \bar{L}_x+\overline{\widetilde{H}}^{\,u}_k,\qquad \widetilde{N}_1\to \widetilde{L}_x+ H^u_k,
\label{lg16}
\end{equation}
where index $x$ changes from $1$ to $4$. At the tree level the rates of these decay modes of $N_1$ and $\widetilde{N}_1$
are set by the Yukawa couplings $h^N_{kx1}$. Supersymmetry implies that
\begin{equation}
\Gamma^{k}_{N_1 \ell_x}+\Gamma^{k}_{N_1 \bar{\ell}_x}=\Gamma^{k}_{N_1 \widetilde{\ell}_x}+\Gamma^{k}_{N_1 \widetilde{\ell}^{*}_x}=
\Gamma^{k}_{\widetilde{N}_1^{*}\ell_x}=\Gamma^{k}_{\widetilde{N}_1 \bar{\ell}_x}=\Gamma^{k}_{\widetilde{N}_1 \widetilde{\ell}_x}=
\Gamma^{k}_{\widetilde{N}_1^{*} \widetilde{\ell}_x^{*}}=\dfrac{|h^N_{kx1}|^2}{8\pi}\,M_1\,,
\label{lg27}
\end{equation}
where superscript $k$ represents either Higgs (Higgsino) if $k=3$ or inert Higgs (inert Higgsino) field if $k=1,\,2$ in the final state.
Here and further we work in a field basis where the charged lepton Yukawa matrix and mass matrix of the right--handed neutrinos
are diagonal. We also assume that supersymmetry breaking scale is negligibly small as compared with $M_1$. As a consequence
all soft SUSY breaking terms can be safely ignored in our calculations of decay asymmetries and rates.

Each decay channel (\ref{lg16}) gives rise to the CP asymmetry that contributes to the generation of total lepton asymmetry.
In the considered case the definition of the CP asymmetries (\ref{lg6}) coming from the decays of the lightest right--handed
neutrino can be generalised in the following way
\begin{equation}
\begin{array}{rcl}
\varepsilon^{k}_{1,\,f}&=&\dfrac{\Gamma^{k}_{N_1 f}-\Gamma^{k}_{N_1 \bar{f}}}
{\sum_{m,\,f'} \left(\Gamma^{m}_{N_1 f'}+\Gamma^{m}_{N_1 \bar{f}'}\right)}\,,
\end{array}
\label{lg29}
\end{equation}
where $f$ and $f'$ may be either $\ell_x$ or $\widetilde{\ell}_x$ while $\bar{f}$ and $\bar{f}'$ should be associated
with either $\bar{\ell}_x$ or $\widetilde{\ell}_x^{*}$. Here $\varepsilon^{3}_{1,\,\ell_n}$ and $\varepsilon^{3}_{1,\,\widetilde{\ell}_n}$
(n=1,2,3) are flavour CP asymmetries that stem from the decays of the lightest right--handed neutrino into leptons (sleptons) and Higgs
doublet $H_u$ (Higgsino $\widetilde{H}_u$) while $\varepsilon^{3}_{1,\,\ell_4}$, $\varepsilon^{3}_{1,\,\widetilde{\ell}_4}$,
$\varepsilon^{1}_{1,\,f}$ and $\varepsilon^{2}_{1,\,f}$ are extra CP asymmetries caused by the new decay channels of $N_1$.
The denominators of Eqs.~(\ref{lg29}) contain a sum of partial decay widths of the lightest right--handed neutrino. For
$\varepsilon^{k}_{1,\,\ell_x}$ this sum includes all possible partial widths of the decays of $N_1$ whose final state involves
leptons and fermion components of $L_4$. The expressions for $\varepsilon^{k}_{1,\,\widetilde{\ell}_x}$ contain in the denominator
a sum of partial decay widths of $N_1$ over all possible decay modes that have either slepton or scalar components of $L_4$
in the final state. The CP asymmetries caused by the decays of the lightest right--handed sneutrino $\varepsilon^{k}_{\widetilde{1},\,f}$
can be defined similarly to the neutrino ones. In this case the right--handed neutrino field in Eqs.~(\ref{lg29}) ought to be replaced
by either $\widetilde{N}_1$ or $\widetilde{N}_1^{*}$.

\begin{figure}
\begin{center}
\includegraphics[width=50mm]{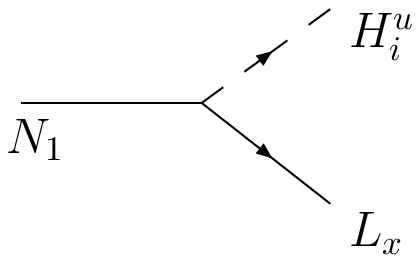}\\
\includegraphics[width=160mm]{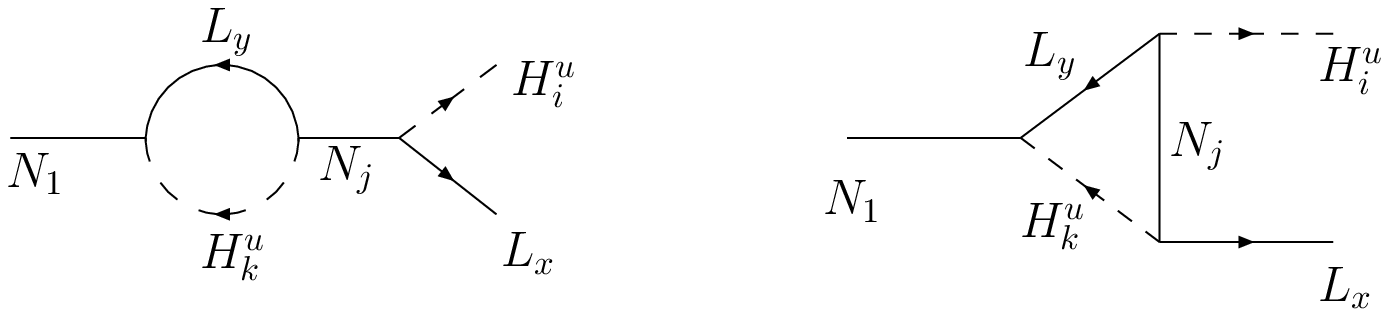}\\
\includegraphics[width=160mm]{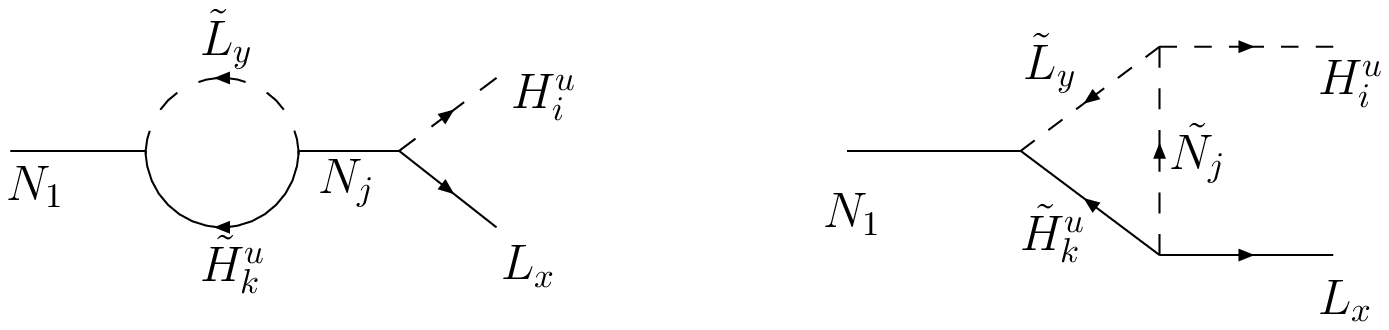}\\
\includegraphics[width=160mm]{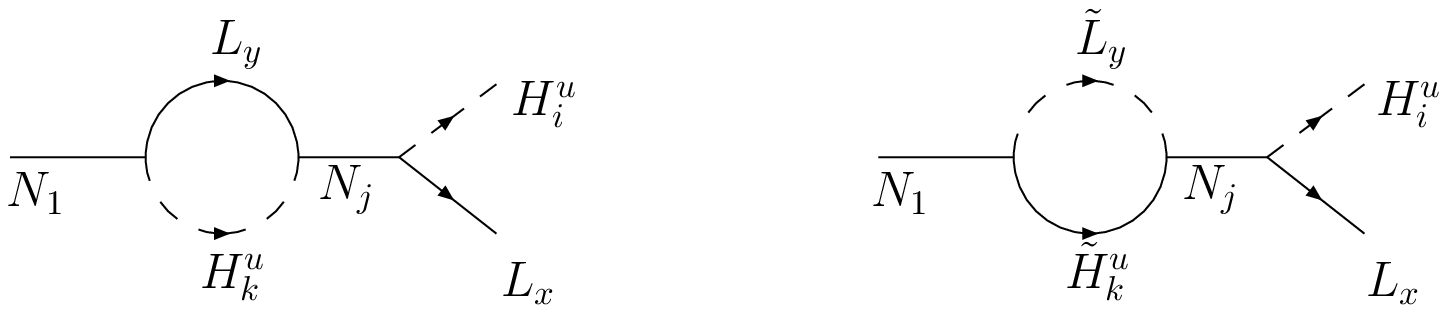}
\caption{Diagrams that give contribution to the CP asymmetries in the E$_6$SSM Model I,
including the presence of two extra inert Higgs doublets, and the fourth
family lepton doublet.}
\end{center}
\end{figure}

As in the SM and MSSM the CP asymmetries within the E$_6$SSM Model I arise due to the interference between the tree--level amplitudes
of the lightest right--handed neutrino decays and the vertex and self--energy corrections to them. The corresponding tree--level and
one--loop diagrams are shown in Fig.~1. After the calculation of one--loop diagrams we get
\begin{equation}
\begin{array}{rcl}
\varepsilon^{k}_{1,\,\ell_x}&=&\varepsilon^{k}_{1,\,\widetilde{\ell}_x}=\varepsilon^{k}_{\widetilde{1},\,\ell_x}=
\varepsilon^{k}_{\widetilde{1},\,\widetilde{\ell}_x}=\dfrac{1}{4\pi A_1}\sum_{j=2,3}\mbox{Im}\biggl\{
A_j h^{N*}_{kx1} h^{N}_{kxj} f^S\left(\dfrac{M^2_j}{M_1^2}\right)\\[3mm]
&+&\sum_{m,\,y} h^{N*}_{my1} h^{N}_{mxj}  h^{N}_{kyj} h^{N*}_{kx1}\,
f^V\left(\dfrac{M^2_j}{M_1^2}\right)\biggr\}\,,
\end{array}
\label{lg30}
\end{equation}
where
$$
\begin{array}{c}
A_j=\sum_{m,y} \left( h^{N*}_{my1} h^{N}_{myj} + \dfrac{M_1}{M_j} h^{N}_{my1} h^{N*}_{myj} \right)\,,\\[3mm]
f^S(z)=\dfrac{2\sqrt{z}}{1-z}\,,\qquad\qquad f^V(z)=-\sqrt{z}\,\ln\left(\dfrac{1+z}{z}\right)\,,
\end{array}
$$
with $k,m=1,2,3$ and $x,y=1,2,3,4$. The terms in the right--hand side of Eq.~(\ref{lg30}) which are proportional to $A_j$ are induced by the
self--energy diagrams while all other terms come from vertex corrections. It is worth to notice here that the coefficients in front
of $f^S(x)$ and $f^V(x)$ are not the same, in contrast to the simplest realisations of Fukugita--Yanagida mechanism in the SM and MSSM.
It means that in general vertex and self--energy contributions to $\varepsilon_{1,\,f}$ and $\varepsilon_{\widetilde{1},\,f}$ are not related
to each other in the considered model. This is a common feature of the models in which right-handed Majorana neutrinos interact with a few
lepton doublets and with a few doublets that have quantum numbers of Higgs fields.

Because inert Higgs and inert Higgsino fields do not carry any lepton number it is convenient to define the overall
CP asymmetries which are associated with each flavour, i.e.
\begin{equation}
\varepsilon^{tot}_{1,\,f}=\sum_{k} \varepsilon^{k}_{1,\,f}\,,\qquad\qquad
\varepsilon^{tot}_{\widetilde{1},\,f}=\sum_{k} \varepsilon^{k}_{\widetilde{1},\,f}\,.
\label{lg32}
\end{equation}
These overall decay asymmetries enter in the right--hand side of Boltzmann equations that describe the evolution of
lepton number densities. The CP asymmetries (\ref{lg32}) can be written in a compact form
\begin{equation}
\varepsilon^{tot}_{1,\,f}=\varepsilon^{tot}_{\widetilde{1},\,f}=
\dfrac{1}{8\pi (\mbox{Tr} \Pi^1)}\sum_{j=2,3}\mbox{Im}\biggl\{A_j \Pi^{j}_{ff} f^S\left(\dfrac{M^2_j}{M_1^2}\right)+
(\Pi^{j})^2_{ff} f^V\left(\dfrac{M^2_j}{M_1^2}\right)\biggr\}\,,
\label{lg33}
\end{equation}
where
\begin{equation}
\Pi^j_{\ell_y \ell_x}=\Pi^j_{\tilde{\ell}_y \tilde{\ell}_x}=\sum_{m} h^{N*}_{my1}h^{N}_{mxj}
\label{lg34}
\end{equation}
are three $4\times 4$ matrices and $A_j= \mbox{Tr} \Pi^{j} + \dfrac{M_1}{M_j} \mbox{Tr} \Pi^{j*}$\,.
Eqs.~(\ref{lg33})--(\ref{lg34}) indicate that despite
a large number of new couplings appeared due to the breakdown of $Z^2_H$ symmetry only some combinations of them contribute
to the generation of lepton asymmetries. The parametrisation of the overall flavour CP asymmetries presented above can be used
in any model in which lightest right-handed neutrino can decay into a few lepton multiplets and a few $SU(2)_W$ doublets
that have quantum numbers of Higgs fields.

In the case of unbroken $Z_2^{H}$ symmetry the analytical expressions for the decay asymmetries (\ref{lg30}) and (\ref{lg33})
are simplified dramatically. In particular, CP asymmetries $\varepsilon^{1}_{1,\,f}$ and $\varepsilon^{2}_{1,\,f}$ which are
associated with the decays of $N_1$ into either scalar or fermion component of inert Higgs superfields $H_{2\alpha}$ vanish
when $Z_2^{H}$ symmetry violating Yukawa couplings tend to zero. The analytical expressions for other decay asymmetries reduce to
\begin{equation}
\begin{array}{c}
\varepsilon^{3}_{1,\,\ell_x}=\varepsilon^{3}_{1,\,\widetilde{\ell}_x}=\varepsilon^{3}_{\widetilde{1},\,\ell_x}=
\varepsilon^{3}_{\widetilde{1},\,\widetilde{\ell}_x}=\dfrac{1}{8\pi}\dfrac{\sum_{j=2,3}
\mbox{Im}\biggl[h^{N*}_{3x1} B_{1j} h^{N}_{3xj} \biggr]}{\sum_{y} |h^N_{3y1}|^2}\,,\\[4mm]
B_{1j}=\sum_{y}\left\{h^{N*}_{3y1} h^N_{3yj} g\left(\dfrac{M^2_j}{M_1^2}\right)+
\dfrac{M_1}{M_j}h^{N}_{3y1} h^{N*}_{3yj}f^S\left(\dfrac{M^2_j}{M_1^2}\right)\right \}\,,\\[4mm]
g(z)=f^{V}(z)+f^{S}(z)=\sqrt{z}\biggl[\dfrac{2}{1-z}-\ln\left(\dfrac{1+z}{z}\right)\biggr]\,,
\end{array}
\label{lg19}
\end{equation}
where $x$ and $y$ vary from 1 to 4.
If the second lightest and heaviest right--handed neutrinos are significantly heavier than the lightest
one, i.e. $M_2,\,M_3 \gg M_1$, the formulae for the CP asymmetries (\ref{lg19}) are simplified even further
\begin{equation}
\begin{array}{c}
\varepsilon^{3}_{1,\,\ell_x}\simeq-\dfrac{3}{8\pi}\sum_{j=2,3}
\dfrac{\mbox{Im}\biggl[(h^{N\dagger} h^{N})_{1j} h^{N*}_{3x1} h^{N}_{3xj}\biggr]}{(h^{N\dagger} h^{N})_{11}}\,\dfrac{M_1}{M_j}\,,
\end{array}
\label{lg21}
\end{equation}
where $(h^{N\dagger} h^{N})_{1j}=\sum_{y} h^{N*}_{3y1} h^{N}_{3yj}$.
From Eq.~(\ref{lg19}) one can see that the self--energy contribution to the flavour CP asymmetries
is twice larger than the vertex one in the considered case.

The derived analytical expressions for the CP asymmetries (\ref{lg19})--(\ref{lg21}) are very similar to
the MSSM ones. Moreover when Yukawa couplings $h^{N}_{34j}\to 0$ extra CP asymmetries induced by the decays
\begin{equation}
N_1\to L_4+H_u,\qquad N_1\to \widetilde{L}_4+\widetilde{H}_u,\qquad
\widetilde{N}_1\to \bar{L}_4+\overline{\widetilde{H}}_u,\qquad \widetilde{N}_1\to \widetilde{L}_4+ H_u,
\label{lg201}
\end{equation}
go to zero and the results for the flavour lepton decay asymmetries obtained within the MSSM are
reproduced. However if Yukawa couplings $h^{N}_{34j}$ have non--zero values the process of generation
of lepton asymmetry in the MSSM and E$_6$SSM with unbroken $Z_2^{H}$ can be entirely different because of
the presence of superfields $L_4$ in the particle spectrum of the E$_6$SSM. Indeed, since $h_{34j}^N$
can be either of the order of or even larger than the Yukawa couplings of the ordinary lepton superfields
to the Higgs doublet $H_u$ the decay rates and CP asymmetries associated with the decays (\ref{lg201})
can be substantially bigger than other decay rates and asymmetries. The fermion and scalar components of
the supermultiplet $L_4$ being produced in the decays of the lightest right--handed neutrino and sneutrino
sequentially decay either to the leptons or to the sleptons changing the induced lepton number asymmetries.

\subsection{CP asymmetries for the Model II}

In the E$_6$SSM Model II there are, in addition to the states in Model I,
exotic leptoquarks which carry baryon and lepton numbers simultaneously.
In this case quark--lepton couplings of $\overline{D}_i$ and $D_i$ in the superpotential
do not violate either baryon or lepton $U(1)$ global symmetries so that these interactions are allowed
from the phenomenological point of view. On the other hand these couplings violate $Z_2^{H}$ symmetry
and therefore the corresponding interactions should be rather weak.

The non--zero complex Yukawa couplings of the leptoquarks to the right--handed Majorana neutrinos (see Eq.(\ref{lg5}))
give rise to extra contributions to the CP asymmetries which correspond to different lepton flavours.
These contributions come from the one--loop self--energy diagrams shown in Fig.~2 that contain virtual (possibly exotic)
quarks and squarks. Because Yukawa couplings of the leptoquarks do not induce any one--loop vertex corrections to the
amplitude of the decay of the lightest right--handed neutrino, lepton decay asymmetries can be
described by Eqs.~(\ref{lg30}) in which $A_2$ and $A_3$ should be replaced by $\widetilde{A}_2$
and $\widetilde{A}_3$ where
\begin{equation}
\widetilde{A}_j=A_j+\dfrac{3}{2}\sum_{m,n} \left(g^{N*}_{mn1} g^{N}_{mnj}+\dfrac{M_1}{M_j}g^{N}_{mn1} g^{N*}_{mnj}
\right)\,.
\label{lg35}
\end{equation}

\begin{figure}
\begin{center}
\includegraphics[width=160mm]{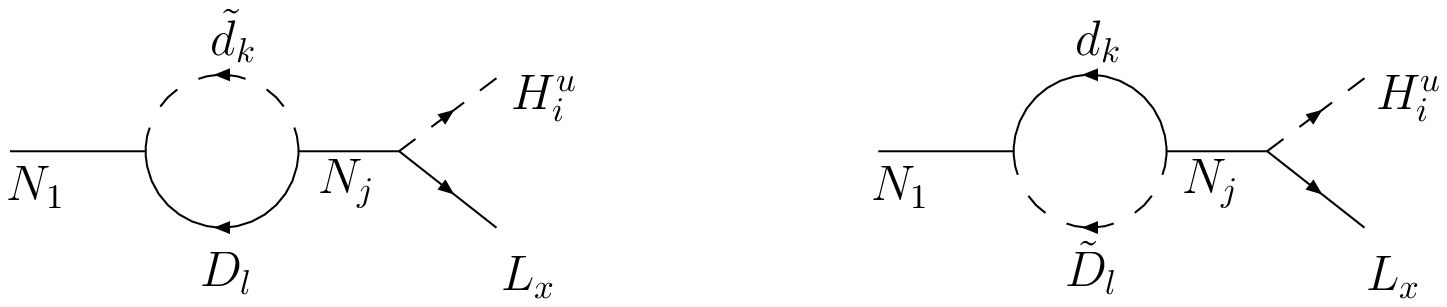}\\
\includegraphics[width=160mm]{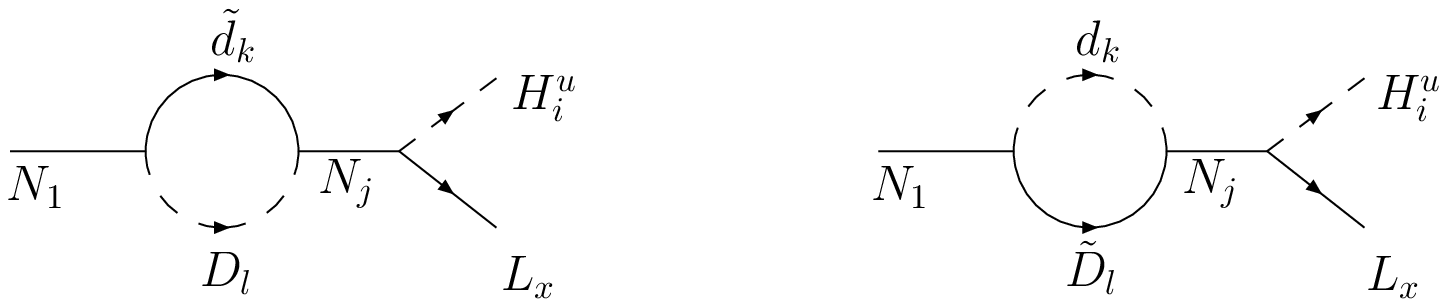}
\caption{Extra one--loop diagrams involving internal leptoquarks $D$
that contribute to the CP asymmetries associated
with the decays $N_1\to L_x +H^u_{k}$ in the E$_6$SSM Model II}
\end{center}
\end{figure}

At the same time the interactions of $D_i$ and $\overline{D}_i$ with $N_1$ and quark superfields
give rise to the new channels of the lightest right--handed neutrino and sneutrino decays
\begin{equation}
N_1\to D_k+\widetilde{d^c}_i,\quad N_1\to \widetilde{D}_k+d^c_i,\quad \widetilde{N}_1\to \overline{D}_k+d_i,\quad
\widetilde{N}_1\to \widetilde{D}_k+\widetilde{d^c}_i,
\label{lg36}
\end{equation}
where $D_k$ and $\widetilde{D}_k$ are fermion and scalar components of leptoquark superfields while
$d_i$ and $\widetilde{d}_i$ are right--handed down type quarks and their superpartners.
When the supersymmetry breaking scale lies considerably lower than the lightest right--handed neutrino mass $M_1$,
the corresponding partial decay widths are determined by the $Z_2^H$ symmetry violating Yukawa couplings
$g^{N}_{ki1}$ only, i.e.
\begin{equation}
\begin{array}{c}
\Gamma^{i}_{N_1 D_k}+\Gamma^{i}_{N_1 \bar{D}_k}=\Gamma^{i}_{N_1 \widetilde{D}_k}+\Gamma^{i}_{N_1 \widetilde{D}^{*}_k}=
\Gamma^{i}_{\widetilde{N}_1^{*}D_k}=\Gamma^{i}_{\widetilde{N}_1 \bar{D}_k}=\\[2mm]
=\Gamma^{i}_{\widetilde{N}_1 \widetilde{D}_k}=\Gamma^{i}_{\widetilde{N}_1^{*} \widetilde{D}_k^{*}}=\dfrac{3|g^{N}_{ki1}|^2}{16\pi}\,M_1\,.
\end{array}
\label{lg37}
\end{equation}

New channels of the decays of the lightest right--handed neutrino (or sneutrino) contribute to the generation of
lepton asymmetry via the sequential decay of leptoquarks and their superpartners at low energies. Due to the lepton number
conservation, each $D_k$ and $\widetilde{D}_k$ produce a lepton in the final state whereas the decay of their antiparticles
leads to the appearance of an antilepton. As a consequence one can calculate lepton CP asymmetries associated with
each additional channel of the lightest right--handed neutrino (or sneutrino) decay (\ref{lg36}). We define the CP asymmetries
caused by the decays of $N_1$ into the exotic quarks (squarks) as follows
\begin{equation}
\begin{array}{c}
\varepsilon^{i}_{1,\,q_k}=\dfrac{\Gamma^{i}_{N_1 q_k}-\Gamma^{i}_{N_1 \bar{q}_k}}
{\sum_{j,\,m} \left(\Gamma^{j}_{N_1 q_m}+\Gamma^{j}_{N_1 \bar{q}_m}\right)}\,.
\end{array}
\label{lg38}
\end{equation}
In Eq.~(\ref{lg38}) $q_k$ can be either leptoquark fermion fields $D_k$ or their scalar superpartners
$\widetilde{D}_k$ whereas $\bar{q}_k$ represents charge conjugate states $\overline{D}_k$ or $\widetilde{D}_k^{*}$.
The superscripts $i$ and $j$ indicate the generation number of the down type quark or its superpartner in the
final state. In the denominator of Eq.~(\ref{lg38}) we sum over possible partial widths of the decays
of $N_1$ either into exotic quark and right--handed down type squark if $\varepsilon^{i}_{1,\,q_k}=\varepsilon^{i}_{1,\,D_k}$
or into exotic squark and ordinary $d$--quark if $\varepsilon^{i}_{1,\,q_k}=\varepsilon^{i}_{1,\,\widetilde{D}_k}$.
The CP asymmetries $\varepsilon^{i}_{\widetilde{1},\,q_k}$ which originate from the decays of the
lightest right--handed sneutrino into the exotic quark (squark) can be defined in a similar way replacing
$N_1$ in Eq.~(\ref{lg38}) by either $\widetilde{N}_1$ or $\widetilde{N}_1^{*}$. It is worth noticing that here
we treat the CP asymmetries for the right--handed neutrino (sneutrino) decays to leptons and leptoquarks
separately. In other words we do not combine together all possible partial widths of the decays
of $N_1$ into exotic quarks (squark) and leptons (sleptons) in the denominator of Eq.~(\ref{lg38}) because
leptoquarks and lepton fields carry different quantum numbers.

\begin{figure}
\begin{center}
\includegraphics[width=50mm]{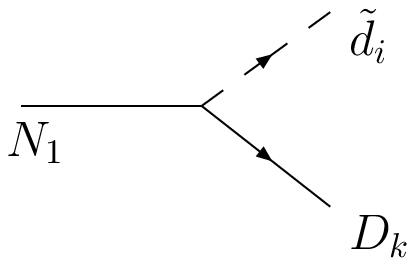}\\
\includegraphics[width=160mm]{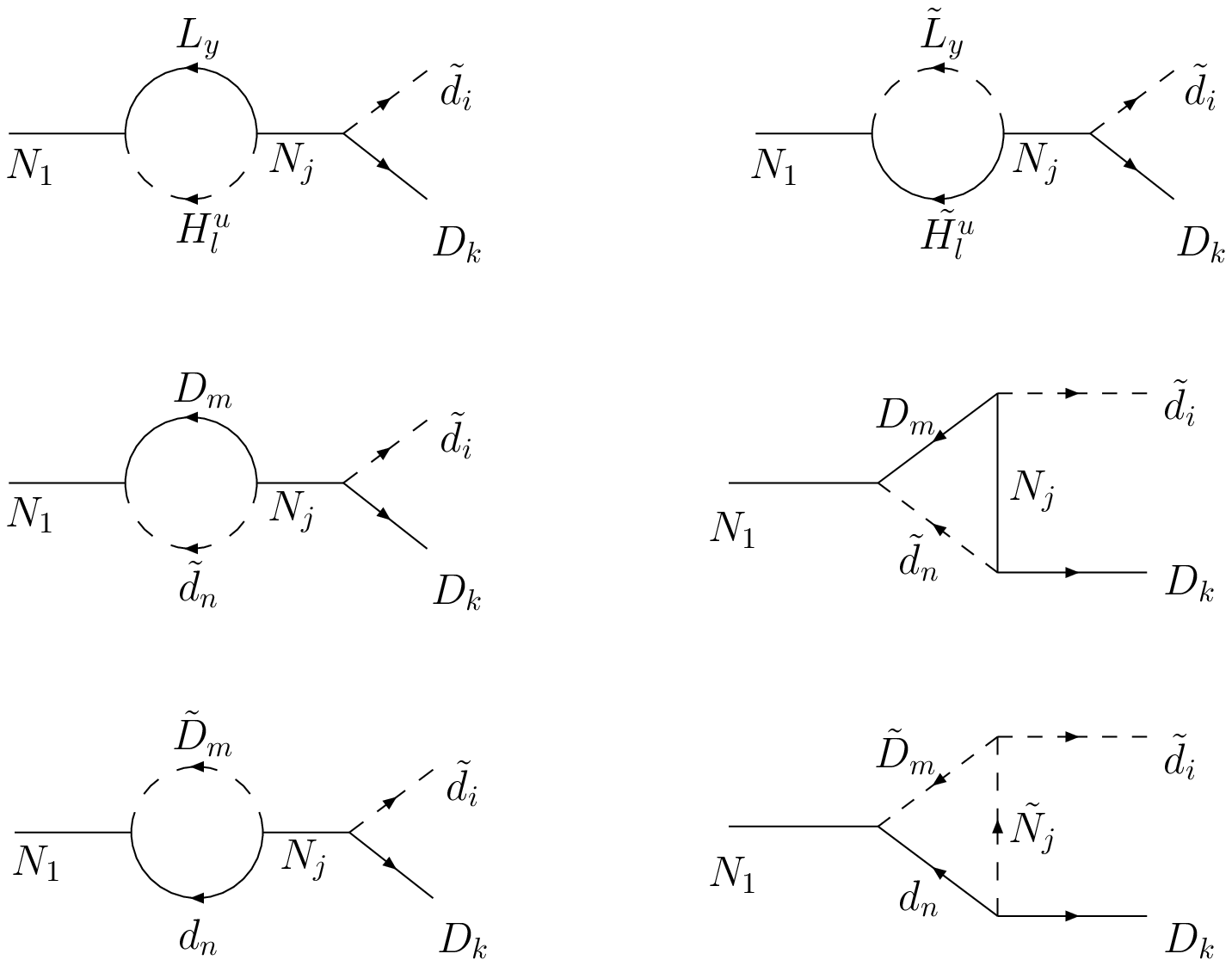}\\
\includegraphics[width=160mm]{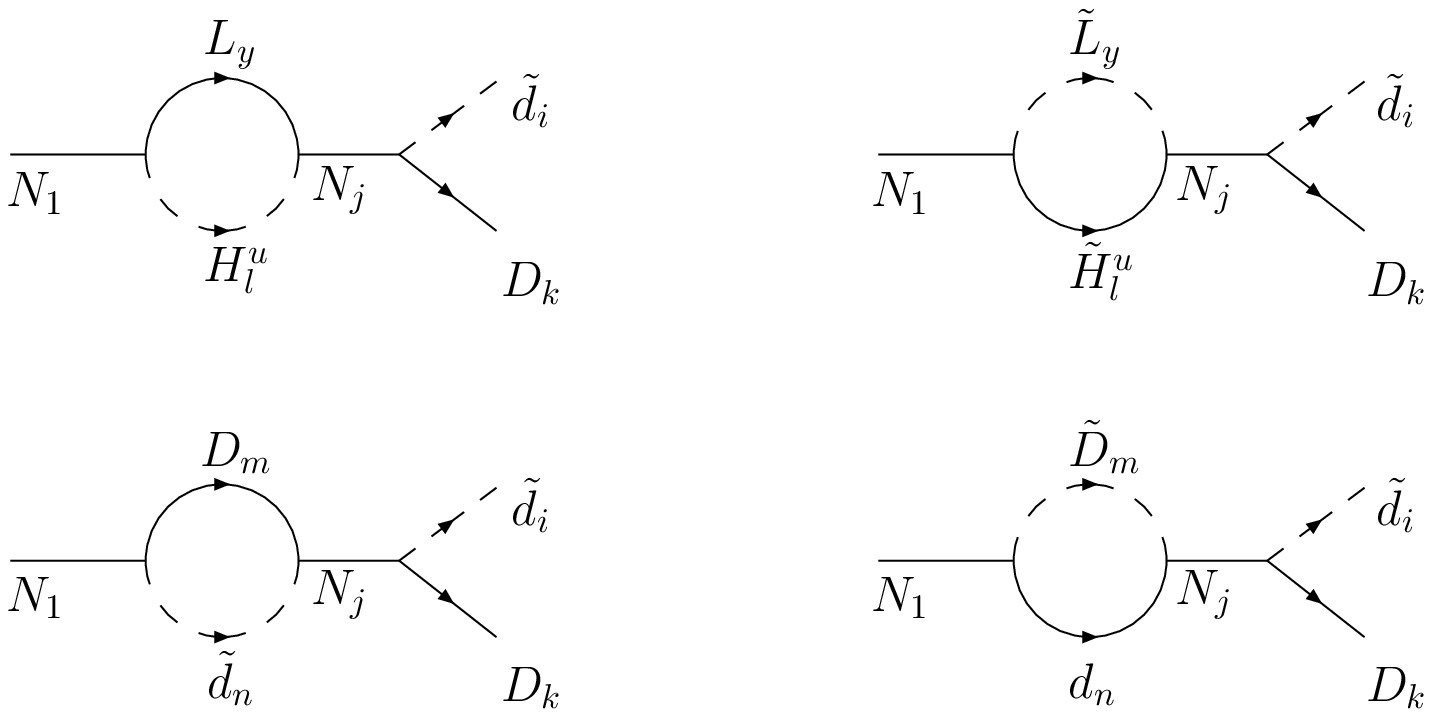}
\caption{Tree--level and one--loop diagrams that give contribution to the CP asymmetries associated with the decays
$N_1\to D_k+d_i$ involving final state leptoquarks $D$
in the E$_6$SSM Model II.}
\end{center}
\end{figure}

In the tree level approximation, the CP asymmetries which are associated with the new decay modes of $N_1$ and
$\widetilde{N}_1$ (\ref{lg36}) vanish. The non--zero values of $\varepsilon^{i}_{1,\,q_k}$ are induced after the
inclusion of one--loop vertex and self--energy corrections to the decay amplitudes of $N_1$ and $\widetilde{N}_1$
if some of the Yukawa couplings of the right--handed Majorana neutrinos to leptons and quarks are complex.
The tree--level and one--loop diagrams that contribute to the decay asymmetries (\ref{lg38}) are presented in Fig.~3.
The interference of the corresponding tree--level decay amplitude with the one--loop corrections yields
\begin{equation}
\begin{array}{rcl}
\varepsilon^{i}_{1,\,D_k}&=&\varepsilon^{i}_{1,\,\widetilde{D}_k}=\varepsilon^{i}_{\widetilde{1},\,D_k}=
\varepsilon^{i}_{\widetilde{1},\,\widetilde{D}_k}=\dfrac{1}{8\pi A_0}\sum_{j=2,3}\mbox{Im}\biggl\{
\widetilde{A}_j g^{N}_{kij} g^{N*}_{ki1} f^S\left(\dfrac{M^2_j}{M_1^2}\right)\\[3mm]
&+& \sum_{m,\,n} g^{N*}_{mn1} g^{N}_{mij} g^{N}_{knj} g^{N*}_{ki1}
f^V\left(\dfrac{M^2_j}{M_1^2}\right)\biggr\}\,,
\end{array}
\label{lg39}
\end{equation}
where $A_0=\sum_{k,\,i}g^{N}_{ki1} g^{N*}_{ki1}$. As before, supersymmetry ensures that the CP asymmetries
originating from the decays of the lightest right--handed neutrino and sneutrino are equal. As in the
case of the lepton decay asymmetries (\ref{lg30}) the terms in the right--hand side of Eqs.~(\ref{lg39})
involving $\widetilde{A}_j$ stem from the self--energy diagrams while all other terms represent vertex
corrections. Again the coefficients in front of $f^S(x)$ and $f^V(x)$ are not equal unlike the simplest
realisations of Fukugita--Yanagida mechanism. From Eq.~(\ref{lg39}) it follows that the decay asymmetries
induced by the additional decay modes (\ref{lg36}) depend not only on the Yukawa couplings of exotic quarks
and squarks to the right--handed neutrino but also on the couplings of the right--handed neutrino to leptons
and sleptons. Extra CP asymmetries (\ref{lg39}) tend to zero when the $Z_2^H$ symmetry violating Yukawa
couplings $g^{N}_{kij}$ vanish.

We can also define the overall decay asymmetries which are associated with each generation of exotic
quarks, i.e.
\begin{equation}
\varepsilon^{tot}_{1,\,q_k}=\sum_{i} \varepsilon^{i}_{1,\,q_k}\,,\qquad\qquad
\varepsilon^{tot}_{\widetilde{1},\,q_k}=\sum_{i} \varepsilon^{i}_{\widetilde{1},\,q_k}\,.
\label{lg40}
\end{equation}
The overall decay asymmetries that stem from the decays of the lightest right--handed neutrino and sneutrino
can be presented in the following form
\begin{equation}
\begin{array}{c}
\varepsilon^{tot}_{1,\,f}=
\dfrac{1}{8\pi (\mbox{Tr} \Pi^1)}\sum_{j=2,3}\mbox{Im}\biggl\{
\widetilde{A}_j
\Pi^{j}_{ff} f^S\left(\dfrac{M^2_j}{M_1^2}\right)+
(\Pi^{j})^2_{ff} f^V\left(\dfrac{M^2_j}{M_1^2}\right)\biggr\}\,,\\[3mm]
\varepsilon^{tot}_{1,\,k}=
\dfrac{1}{8\pi (\mbox{Tr} \Omega^1)}\sum_{j=2,3}\mbox{Im}\biggl\{
\widetilde{A}_j
\Omega^{j}_{kk} f^S\left(\dfrac{M^2_j}{M_1^2}\right)+
(\Omega^{j})^2_{kk} f^V\left(\dfrac{M^2_j}{M_1^2}\right)\biggr\}\,,\\[4mm]
\widetilde{A}_j=\mbox{Tr} \Pi^{j}+ \dfrac{M_1}{M_j} \mbox{Tr} \Pi^{j*}+
\dfrac{3}{2} \left( \mbox{Tr} \Omega^{j}+ \dfrac{M_1}{M_j} \mbox{Tr} \Omega^{j*}\right)\,,
\end{array}
\label{lg41}
\end{equation}
where we set $\varepsilon^{tot}_{1,\,D_k}=\varepsilon^{tot}_{1,\,\widetilde{D}_k}=
\varepsilon^{tot}_{\widetilde{1},\,D_k}=\varepsilon^{tot}_{\widetilde{1},\,\widetilde{D}_k}=\varepsilon^{tot}_{1,\,k}$,
$\Omega^{j}_{ki}=\sum_{m} g^{N*}_{km1} g^{N}_{imj}$ while $\Pi^j_{mn}$ are given by Eqs.~(\ref{lg34}).
Compact parametrisation of the overall CP asymmetries (\ref{lg41}) allows elimination of a number of parameters
on which total lepton asymmetry does not depend.

\section{Numerical Results and Discussions}

We now consider the impact of new particles and interactions appearing in the E$_6$SSM on the numerical
values of the lepton CP asymmetries originating from the decays of the lightest right--handed neutrino
and sneutrino. These decay asymmetries depend on all Yukawa couplings of neutrino superfields. Because
the purpose of our studies here is to reveal the impact of extra couplings on the CP asymmetries we shall
fix the Yukawa couplings of the lightest right--handed neutrino and sneutrino to lepton and Higgs superfields
so that the observed pattern of neutrino masses and mixing angles is reproduced.
Here, as an example, we concentrate on the see--saw models \cite{see-saw} with sequential dominance (SD) of
right--handed neutrinos \cite{SD}--\cite{King:2002qh} which lead to the appropriate neutrino spectrum in a technically
natural way, i.e. small perturbations in the high energy input parameters do not change substantially the neutrino mass
splittings at low energies. This means that small neutrino mass splittings are preserved in the presence of radiative
corrections\footnote{In general the radiative corrections in see--saw models may be sufficient to destroy (or create)
the cancellations necessary to achieve the desired mass hierarchy \cite{RadCor}.}.

\subsection{Constrained Sequential Dominance}

To review how sequential dominance works we begin by writing the right--handed neutrino Majorana mass
matrix in a diagonal basis as
\begin{equation}
M_{RR}=\left(
\begin{array}{ccc}
M_1 & 0   & 0\\
0   & M_2 & 0\\
0   & 0   & M_3
\end{array}
\right)
\label{lg42}
\end{equation}
and the matrix of Yukawa couplings of the right--handed neutrino to lepton and Higgs fields $h^N_{ij}$
in terms of $(1,\,3)$ column vectors $A_i$, $B_i$ and $C_i$ as
\begin{equation}
h_{ij}^N=(A\quad B\quad C)=\left(
\begin{array}{ccc}
d & a & a'\\
e & b & b'\\
f & c & c'
\end{array}
\right).
\label{lg43}
\end{equation}
As before we assume that $M_1\ll M_2\ll M_3$. We shall also assume that $|d|\ll |e|\sim |f|$. The breakdown of electroweak
symmetry induces Majorana mass terms for the left--handed neutrino via the Yukawa interactions of the neutrino
with the Higgs fields. After the integrating out the right--handed neutrino we get
\begin{equation}
\mathcal{L}^{\nu}_{mass}=\dfrac{(\nu_i^T A_i)(A_j^T \nu_j)}{M_1}v_2^2+\dfrac{(\nu_i^T B_i)(B_j^T \nu_j)}{M_2}v_2^2+
\dfrac{(\nu_i^T C_i)(C_j^T \nu_j)}{M_3}v_2^2\,,
\label{lg44}
\end{equation}
where $v_2$ is a VEV of the Higgs doublet $H_u$. The $3\times 3$ mass matrix of the left--handed neutrino
induced by $\mathcal{L}^{\nu}_{mass}$ can be diagonalised by means of a unitary transformation that may be
written as a sequence of transformations
\begin{equation}
V^{\nu\,\dagger}=P\,R_{23} U_{13} R_{12} P_{12}\,,
\label{lg45}
\end{equation}
where
\begin{equation}
\begin{array}{rclrcl}
P&=&\left(
\begin{array}{ccc}
e^{i\omega_1} & 0             & 0\\
0             & e^{i\omega_2} & 0\\
0             & 0             &  e^{i\omega_3}
\end{array}
\right)\,,&\qquad
R_{23}&=&\left(
\begin{array}{ccc}
1 & 0      & 0\\
0 & c^{\nu}_{23} & s^{\nu}_{23}\\
0 & -s^{\nu}_{23}& c^{\nu}_{23}
\end{array}
\right)\,,\\[0.4cm]
U_{13}&=&\left(
\begin{array}{ccc}
c^{\nu}_{13}                    & 0 & s^{\nu}_{13}\,e^{-i\delta^{\nu}}\\
0                               & 1 & 0\\
-s^{\nu}_{13}\,e^{i\delta^{\nu}}& 0 & c^{\nu}_{13}
\end{array}
\right)\,,&\qquad
R_{12}&=&\left(
\begin{array}{ccc}
c^{\nu}_{12}  & s^{\nu}_{12} & 0\\
-s^{\nu}_{12} & c^{\nu}_{12} & 0\\
0             & 0            & 1
\end{array}
\right)\,,\\[0.4cm]
P_{12}&=&\left(
\begin{array}{ccc}
e^{i\beta_1} & 0             & 0\\
0             & e^{i\beta_2} & 0\\
0             & 0            & 1
\end{array}
\right)\,,&&&
\end{array}
\label{lg46}
\end{equation}
and $s^{\nu}_{ij}=\sin\theta^{\nu}_{ij}$, $c^{\nu}_{ij}=\cos\theta^{\nu}_{ij}$.
The phase matrix $P$ in the right hand side of Eq.~(\ref{lg45}) may always be removed by an
additional charged lepton phase rotation.

The Pontecorvo--Maki--Nakagawa--Sakata (PMNS) neutrino mixing matrix $U_{PMNS}$ \cite{PMNS} is a product of
unitary matrices $V^{E}$ and $V^{\nu\,\dagger}$, where $V^{E}$ is associated with the diagonalisation
of the charged lepton mass matrix. Since the charged lepton mixing angles are expected to be
small $U_{PMNS}\approx V^{\nu\,\dagger}$ in the first approximation. The only exception is
$\theta_{13}$. The CHOOZ experiment sets a stringent constraint on the value of $\theta_{13}\lesssim 0.2$
\cite{Apollonio:1999ae}. Because $\theta_{13}$ is small it receives important contributions not just from
$\theta^{\nu}_{13}$, but also from the charged lepton angles \cite{King:2002nf}. Further we will assume
that $\theta^{\nu}_{13}\ll 1$.

The sequential dominance implies that the first term in Eq.~(\ref{lg44}) gives a dominant contribution
to the mass matrix of the left--handed neutrino, the second term is subdominant whereas the contribution
of the last term in Eq.~(\ref{lg44}) is negligible \cite{SD}--\cite{King:2002qh}. This structure of the mass
terms guarantees that the mass of the heaviest left--handed neutrino $m_3$ is much larger than the mass of
the second lightest one. If the heaviest left--handed neutrino is denoted $\nu_3$ then sequential dominance
results in the physical neutrino eigenstate $\nu_3\simeq d\,\nu_e+e\,\nu_{\mu}+f\,\nu_{\tau}$ with the mass \cite{King:2002nf}
\begin{equation}
|m_3|\simeq (|d|^2+|e|^2+|f|^2)v_2^2/M_1\,.
\label{lg47}
\end{equation}
Two other orthogonal combinations of neutrinos remain massless in the leading approximation. The requirement
of a small angle $\theta^{\nu}_{13}$ implies that $|d|\ll |e|, |f|$. Then the atmospheric angle $\theta_{23}$
is given by \cite{King:2002nf}
\begin{equation}
\tan\theta_{23}\approx \tan\theta^{\nu}_{23} \approx \frac{|e|}{|f|}\,.
\label{lg48}
\end{equation}

Although the leading approximation allows us to get an appropriate description of atmospheric neutrino data,
we need to go beyond it to account for the data of other neutrino experiments. The contribution of the sub--leading
right--handed neutrino does not substantially change the mass of the heaviest left--handed neutrino state (\ref{lg47})
and atmospheric angle (\ref{lg48}). However it gives rise to non--zero second lightest neutrino mass.
The sub--leading contributions to the left--handed neutrino mass matrix also induce mixing between the heaviest
and other left--handed neutrino states. The neutrino mass matrix can be reduced to the block diagonal form
by means of unitary transformations $U_{13}$ if
\begin{equation}
\theta^{\nu}_{13}\approx e^{i(\tilde{\phi}+\phi_{a}-\phi_e)}\dfrac{|a|(e^{*}b+f^{*}c)}{(|e|^2+|f|^2)^{3/2}}\dfrac{M_1}{M_2}
+ e^{i(\tilde{\phi}+\phi_{d}-\phi_e)}\dfrac{|d|}{\sqrt{|e|^2+|f|^2}}\,,
\label{lg49}
\end{equation}
where $\phi_x$ are the phases of Yukawa couplings, i.e. $x=|x|e^{i\phi_{x}}$. The relative phase $\phi_e-\phi_f$ is chosen so
that the angle $\theta^{\nu}_{23}$ is real. The phase $\tilde{\phi}$ is fixed by the requirement that the angle $\theta_{13}$
is real and positive. When $d=0$ we get
\begin{equation}
\tilde{\phi}=\phi_e-\phi_a-\zeta\,, \qquad\qquad \zeta=\arg(e^{*}b+f^{*}c)\,.
\label{lg50}
\end{equation}
It is worth to notice here that the angle $\theta^{\nu}_{13}$ is automatically small in the considered approximation.

Finally, the left--handed neutrino mass matrix can be completely diagonalised by the $R_{12}$ rotation. Then the second lightest
left--handed neutrino gets mass \cite{King:2002nf}
\begin{equation}
|m_2|\simeq \dfrac{|a|^2 v_2^2}{M_2 \sin^2 \theta^{\nu}_{12}}\,,
\label{lg51}
\end{equation}
while the solar angle is given by \cite{King:2002nf}
\begin{equation}
\begin{array}{c}
\tan\theta_{12}\approx\tan\theta^{\nu}_{12}\simeq\dfrac{a}{b\cos\theta_{23}-c e^{i(\phi_e-\phi_f)}\sin\theta_{23}}=
\dfrac{|a|}{|b| c_{23}\cos\phi'_b-|c| s_{23}\cos{\phi'_c}}\,,\\[4mm]
\phi'_{b}=\phi_b-\phi_a-\tilde{\phi}-\delta\,,\qquad \phi'_{c}=\phi_c-\phi_a+\phi_e-\phi_f-\tilde{\phi}-\delta\,.
\end{array}
\label{lg52}
\end{equation}
Once again the phases can be chosen so that $\tan\theta^{\nu}_{12}$ is real and positive. This can be achieved if
phases $\phi'_b$ and $\phi'_c$ satisfy the condition
\begin{equation}
|b| c_{23}\sin\phi'_b\approx |c| s_{23}\sin{\phi'_c}\,.
\label{lg53}
\end{equation}
Note that in contrast with $\theta^{\nu}_{13}$ the solar angle (\ref{lg52}) is completely determined by the sub--leading
couplings due to a natural cancellation of the leading contributions. Therefore this angle should be relatively large.
The lightest left--handed neutrino state remains massless in the considered approximation. Its mass is generated by the
sub--sub--leading couplings of the heaviest right--handed neutrino, i.e.
\begin{equation}
|m_1|\simeq O\left(\dfrac{|C|^2 v_2^2}{M_3}\right)\,.
\label{lg54}
\end{equation}
Thus sequential dominance results in a full neutrino mass hierarchy $m_1\ll m_2\ll m_3$. Because SD
does not require any fine tuning the contribution of radiative corrections to the neutrino masses and mixing angles
is expected to be quite small, at the level of a few per cent \cite{King:2000hk}.

Current neutrino oscillation data point strongly to a specific form for the lepton mixing matrix with
effective bimaximal mixing of $\nu_{\mu}$ and $\nu_{\tau}$ at the atmospheric scale and effective trimaximal mixing
for $\nu_e$, $\nu_{\mu}$ and $\nu_{\tau}$ at solar scale (tri--bimaximal mixing \cite{tribimax}). In the tri--bimaximal
mixing scenario the PMNS matrix takes a form
\begin{equation}
U_{PMNS}\simeq\left(
\begin{array}{ccc}
\sqrt{\dfrac{2}{3}}  & \sqrt{\dfrac{1}{3}} & 0\\
-\sqrt{\dfrac{1}{6}} & \sqrt{\dfrac{1}{3}} & \sqrt{\dfrac{1}{2}}\\
\sqrt{\dfrac{1}{6}}  & -\sqrt{\dfrac{1}{3}}& \sqrt{\dfrac{1}{2}}
\end{array}
\right)\,.
\label{lg55}
\end{equation}
Comparing matrix (\ref{lg55}) with the general parametrisation of the neutrino mixing matrix (\ref{lg45}) one can easily
establish that tri--bimaximal mixing scenario corresponds to $\theta_{13}=0$, $\sin{\theta_{12}}=1/\sqrt{3}$ and $\theta_{23}=\pi/4$.
Within the framework of sequential dominance the vanishing of the mixing angle $\theta_{13}$ can be naturally achieved when
\begin{equation}
d\simeq 0\,,\qquad\qquad e^{*}b+f^{*}c=(A^{\dagger} B)\simeq 0\,.
\label{lg56}
\end{equation}
Since in this case the bimaximal mixing between $\nu_{\mu}$ and $\nu_{\tau}$ implies that $|e|=|f|$ the conditions (\ref{lg56}) constrain
the Yukawa couplings of the second lightest right--handed neutrino. In particular, from Eq.~(\ref{lg56}) it follows that $|b|=|c|$. Taking
into account that tri--bimaximal mixing also requires $\sin\theta_{12}=1/\sqrt{3}$ one can show that within the sequential dominance the
Yukawa couplings of the lightest and second lightest right--handed neutrinos which correspond to the tri--bimaximal mixing scenario can be
always chosen so that
\begin{equation}
d\simeq 0\,,\qquad\qquad f=-e=|A|e^{i\phi_A}\,,\qquad\qquad a=b=c=|B|e^{i\phi_B}\,.
\label{lg57}
\end{equation}
This is so--called constrained sequential dominance (CSD) \cite{King:2005bj}. Note that CSD does not constrain the Yukawa couplings of the heaviest
right--handed neutrino $a'$, $b'$ and $c'$ because they only give sub--sub--dominant contribution to the neutrino mass matrix.
Different issues concerning the leptogenesis in the neutrino models based on the seesaw mechanism and sequential right--handed neutrino dominance
were discussed in \cite{King:2002qh}, \cite{Antusch:2006cw}.

\subsection{Results of numerical analysis}

\subsubsection{E$_6$SSM with unbroken $Z^2_H$ symmetry}

With the assumption of the constrained sequential dominance we calculate the values of the decay asymmetries in the E$_6$SSM.
According to CSD one can ignore the contribution of the heaviest right--handed neutrino so that the analytical expressions
for the CP asymmetries derived in Section 3 are considerably simplified. We start our analysis from the E$_6$SSM with exact
$Z^2_H$ symmetry. In this case there is only one extra CP asymmetry associated with the decay of the lightest right--handed
neutrino into scalar (fermion) components of the fourth lepton doublet superfield $L_4$ and Higgsinos (Higgs bosons). Substituting
the pattern of Yukawa couplings that corresponds to the constrained sequential dominance into Eqs.~(\ref{lg21}) and neglecting
the contribution of the heaviest right--handed neutrino to the CP asymmetries we get
\begin{equation}
\begin{array}{rcl}
\varepsilon^3_{1,\,L_4}&\simeq&\dfrac{3}{8\pi}
\dfrac{|h^{N}_{H^u_3 L_4 N_1}|^2 |h^{N}_{H^u_3 L_4 N_2}|^2\sin\phi_{L}}
{2|A|^2+|h^{N}_{H^u_3 L_4 N_1}|^2}\,\dfrac{M_1}{M_2}\,,\qquad\qquad \varepsilon^3_{1,\,e}=0\,,\\[4mm]
\varepsilon^3_{1,\,\tau}&\simeq & -\varepsilon^3_{1,\,\mu}\simeq
\dfrac{3}{8\pi}\dfrac{|h^{N}_{H^u_3 L_4 N_1}||h^{N}_{H^u_3 L_4 N_2}||A||B|\sin\phi_{\mu\tau}}
{2|A|^2+|h^{N}_{H^u_3 L_4 N_1}|^2}\,\dfrac{M_1}{M_2}\,,\\[4mm]
\phi_{\mu\tau}&=&\phi_{41}+\phi_A-\phi_{42}-\phi_B\,,\qquad\qquad\quad
\phi_{L}=2(\phi_{41}-\phi_{42})\,,
\end{array}
\label{lg61}
\end{equation}
where $h^{N}_{H^u_3 L_4 N_1}\equiv h^{N}_{341}=h^{N}_{41}$, $h^{N}_{H^u_3 L_4 N_2}\equiv h^{N}_{342}=h^{N}_{42}$,
$h^{N}_{H^u_3 L_4 N_1}=|h^{N}_{H^u_3 L_4 N_1}| e^{i\phi_{41}}$ and $h^{N}_{H^u_3 L_4 N_2}=|h^{N}_{H^u_3 L_4 N_2}| e^{i\phi_{42}}$.
Note that in the limit when $h^{N}_{H^u_3 L_4 N_1}$ and $h^{N}_{H^u_3 L_4 N_2}$ go to zero all CP asymmetries vanish. This is not
an accident. When Yukawa couplings $h^{N}_{H^u_3 L_4 N_1}$ and $h^{N}_{H^u_3 L_4 N_2}$ tend to zero the interactions of
the right--handed neutrinos with the Higgs and lepton superfields are exactly the same as in the MSSM. At the same time
the conditions (\ref{lg56}) which result in the natural realisation of the tri--bimaximal mixing scenario in the framework
of sequential dominance ensure the vanishing of all decay asymmetries within the SM and the MSSM. Thus the induced values of the
lepton decay asymmetries (\ref{lg61}) are entirely caused by the new particles and interactions appearing in the E$_6$SSM.

The CP asymmetries (\ref{lg61}) also vanish when all Yukawa couplings are real, i.e. CP invariance
in the lepton sector is preserved. The decay asymmetries $\varepsilon^3_{1,\,L_4}$ and $\varepsilon^3_{1,\,\tau}=-\varepsilon^3_{1,\,\mu}$
attain their maximum absolute values when $\sin\phi_{L}$ and $\sin\phi_{\mu\tau}$ are equal to $\pm 1$ respectively.
The maximum absolute values of the CP asymmetries (\ref{lg61}) are given by
\begin{equation}
\begin{array}{rcl}
|\varepsilon^3_{1,\,L_4}| &\simeq & \dfrac{3}{8\pi}
\dfrac{|h^{N}_{H^u_3 L_4 N_1}|^2 |h^{N}_{H^u_3 L_4 N_2}|^2}{2|A|^2+|h^{N}_{H^u_3 L_4 N_1}|^2}\,\dfrac{M_1}{M_2}\,,\\
|\varepsilon^3_{1,\,\tau}| &= & |\varepsilon^3_{1,\,\mu}|\simeq
\dfrac{3}{8\pi}\dfrac{|h^{N}_{H^u_3 L_4 N_1}||h^{N}_{H^u_3 L_4 N_2}||A||B|}{2|A|^2+|h^{N}_{H^u_3 L_4 N_1}|^2}\,\dfrac{M_1}{M_2}\,.
\end{array}
\label{lg62}
\end{equation}
The dependence of the maximum values of $|\varepsilon^3_{1,\,L_4}|$ and $|\varepsilon^3_{1,\,\tau}|=|\varepsilon^3_{1,\,\mu}|$ on the
absolute values of the additional Yukawa couplings $|h^{N}_{H^u_3 L_4 N_1}|$ and $|h^{N}_{H^u_3 L_4 N_2}|$ is examined in Fig.~4\,
where we fix $(M_2/M_1)=10$. To avoid problems related with the overproduction of gravitinos we assume that the mass of the lightest
right--handed neutrino is relatively small $M_1\simeq 10^6\,\mbox{GeV}$. We also set $v_2=v\simeq 246\,\mbox{GeV}$ that
corresponds to large values of $\tan\beta$ and choose parameters $|A|$ and $|B|$ so that the observed neutrino mass--squared
differences are reproduced (see, for example, \cite{Fogli:2005cq}).

In Figs.~4a and 4b the dependence of the maximum value of $|\varepsilon^3_{1,\,\tau}|=|\varepsilon^3_{1,\,\mu}|$ on $|h^{N}_{H^u_3 L_4 N_1}|$
and $|h^{N}_{H^u_3 L_4 N_2}|$ is studied whereas in Figs.~4c and 4d we plot the maximum value of $|\varepsilon^3_{1,\,L_4}|$ as a function
of new Yukawa couplings. From Eqs.~(\ref{lg62}) and Figs.~4a and 4c it follows that both maximum absolute values of the
CP asymmetries (\ref{lg62}) grow monotonically with increasing of $|h^{N}_{H^u_3 L_4 N_2}|$. The dependence of $|\varepsilon^3_{1,\,L_4}|$ and
$|\varepsilon^3_{1,\,\tau}|=|\varepsilon^3_{1,\,\mu}|$ on $|h^{N}_{H^u_3 L_4 N_1}|$ is more complicated. At small values of $|h^{N}_{H^u_3 L_4 N_1}|$
these decay asymmetries are small and increase when $|h^{N}_{H^u_3 L_4 N_1}|$ becomes larger. However if $|h^{N}_{H^u_3 L_4 N_1}|$ is much larger
than $|A|$ the maximum absolute values of $|\varepsilon^3_{1,\,\tau}|=|\varepsilon^3_{1,\,\mu}|$ is inversely proportional to
$|h^{N}_{H^u_3 L_4 N_1}|$ and therefore diminishes with increasing of $|h^{N}_{H^u_3 L_4 N_1}|$ while $|\varepsilon^3_{1,\,L_4}|$
reaches its saturation limit (see Figs.~4b and 4d). The CP asymmetries $|\varepsilon^3_{1,\,\tau}|=|\varepsilon^3_{1,\,\mu}|$ attain their
maximal possible value at $|h^{N}_{H^u_3 L_4 N_1}|\simeq \sqrt{2}|A|$. Thus we establish the following theoretical restrictions on
the values of decay asymmetries
\begin{equation}
|\varepsilon^3_{1,\,L_4}|\lesssim \dfrac{3 M_1}{8\pi M_2} |h^{N}_{H^u_3 L_4 N_2}|^2\,,\qquad\qquad
|\varepsilon^3_{1,\,\tau}|=|\varepsilon^3_{1,\,\mu}|\lesssim \dfrac{3 \sqrt{2} M_1}{32\pi M_2}|h^{N}_{H^u_3 L_4 N_2}||B|\,.
\label{lg63}
\end{equation}
One can easily see that the theoretical upper bounds on the absolute values of the CP asymmetries (\ref{lg63}) are determined
by the Yukawa couplings of the second lightest right--handed neutrino and do not depend on the Yukawa couplings of the lightest
right--handed neutrino. In general the maximal absolute values of decay asymmetries diminish when the couplings $|h^{N}_{H^u_3 L_4 N_1}|$
and $|h^{N}_{H^u_3 L_4 N_2}|$ decrease (see Fig.~5).

There~ is~ also~ another~ general~ tendency~ that~ should~ be~ mentioned~ here. When~ $M_1\ll 10^{13}-10^{14}\,\mbox{GeV}$ the absolute
value of the CP asymmetry associated with the decay $N_1\to L_4+H_{u}$ tend to be considerably larger than lepton decay asymmetries
$\varepsilon^3_{1,\,\mu}$ and $\varepsilon^3_{1,\,\tau}$ (see Figs.~4--5). This happens because lower masses of the right--handed neutrinos
require smaller values of the Yukawa couplings of the Higgs doublet $H_u$ to leptons. Otherwise the observed neutrino mass--squared
differences can not be reproduced within the framework of sequential dominance. From Eqs.~(\ref{lg47}) and (\ref{lg51})
it follows that $|A|\propto \sqrt{M_1|m_3|/v^2}$ while $|B|\propto \sqrt{M_2|m_2|/v^2}$. Thus for a fixed ratio $M_1/M_2$
the maximal possible values of the decay asymmetries $|\varepsilon^3_{1,\,\mu}|$ and $|\varepsilon^3_{1,\,\tau}|$ (\ref{lg63})
diminishes as $\sqrt{M_1}$ when $M_1$ decreases. In fact, the decrease of lepton CP asymmetries with the mass of the lightest
right--handed neutrino is a common feature of most see--saw models. This results in the lower bound on the lightest right--handed
neutrino mass: $M_1\gtrsim 10^9\,\mbox{GeV}$ \cite{lower-bound}. At the same time the results of our analysis presented in Figs.~5
demonstrate that within the E$_6$SSM with unbroken $Z_2^H$ it is possible to generate an appreciable value of the CP asymmetry
$|\varepsilon^3_{1,\,L_4}|=10^{-6}-10^{-4}$ even for $M_1=10^6\,\mbox{GeV}$. This can be achieved if the Yukawa couplings of the
fourth lepton doublet $L_4$ to the Higgs fields $H_u$ vary from $0.01$ to $0.1$. At low energies the induced lepton asymmetry is
transferred to the ordinary lepton asymmetries via the decays of heavy $L_4$ and $\tilde{L}_4$ into leptons (sleptons) and
Higgs fields $H_d$ (Higgsinos $\tilde{H}_d$).

\subsubsection{E$_6$SSM Model I}

In the case of the E$_6$SSM Model I two generations of inert--Higgs superfields $H^u_{\alpha}$ ($\alpha=1,2$) contribute to
$\varepsilon_{1,\,\ell_x}$ through loop diagrams and give rise to a set of extra decay asymmetries
$\varepsilon^{\alpha}_{1,\,\ell_x}$ defined by Eq.~(\ref{lg29}). Because the Yukawa couplings of $H^u_{\alpha}$ to the
quarks and leptons of the first two generation are expected to be rather small in order to avoid non--diagonal flavour transitions
we assume that inert Higgs fields couple to the third generation fermions only. To simplify our analysis further we also assume that
only one inert Higgs doublet $H^u_{2}$ has non--zero couplings with the doublet of leptons of the third generation
and right--handed neutrinos. Then the analytic expression (\ref{lg33}) for the overall CP asymmetries reduces to
\begin{equation}
\begin{array}{rcl}
\varepsilon^{tot}_{1,\,\mu}&\simeq&\dfrac{1}{4\pi}
\dfrac{|h^N_{H^u_2 L_3 N_1}||h^N_{H^u_2 L_3 N_2}||A||B|\sin\phi_{\mu}}
{2|A|^2+|h^N_{H^u_2 L_3 N_1}|^2}\,\dfrac{M_1}{M_2}\,,\qquad\quad \varepsilon^{tot}_{1,\,e}=0\,,\\[4mm]
\varepsilon^{tot}_{1,\,\tau}&\simeq &
\dfrac{\biggl(4|h^N_{H^u_2 L_3 N_1}||h^N_{H^u_2 L_3 N_2}||A||B|\sin\phi_{\mu}+3|h^N_{H^u_2 L_3 N_1}|^2|h^N_{H^u_2 L_3 N_2}|^2\sin\phi_{\tau}\biggr)}
{8\pi(2|A|^2+|h^N_{H^u_2 L_3 N_1}|^2)}\,\dfrac{M_1}{M_2}\,,\\[4mm]
\phi_{\mu}&=&\phi_{231}+\phi_A-\phi_{232}-\phi_B\,,\qquad\qquad\quad \phi_{\tau}=2(\phi_{231}-\phi_{232})\,,
\end{array}
\label{lg64}
\end{equation}
where $h^N_{H^u_2 L_3 N_1}\equiv h^N_{231}$, $h^N_{H^u_2 L_3 N_2}\equiv h^N_{232}$,
$h^N_{H^u_2 L_3 N_1}=|h^N_{H^u_2 L_3 N_1}|e^{i\phi_{231}}$ and \linebreak[4]
$h^N_{H^u_2 L_3 N_2}=|h^N_{H^u_2 L_3 N_2}|e^{i\phi_{232}}$.
Here, to clarify the contribution of the inert--Higgs doublet,
we set all Yukawa couplings of $L_4$ to the right--handed neutrinos to be zero.

As before the overall CP asymmetries (\ref{lg64}) vanish in the MSSM limit of the E$_6$SSM when $h^N_{H^u_2 L_3 N_1}$ and $h^N_{H^u_2 L_3 N_2}$
go to zero. The decay asymmetries (\ref{lg64}) also tend to zero if CP invariance is preserved in the lepton sector, i.e. phases of all Yukawa
couplings vanish. Once again $\varepsilon^{tot}_{1,\,\mu}$ and $\varepsilon^{tot}_{1,\,\tau}$ reach their maximum absolute values when
$\sin\phi_{\mu}$ and $\sin\phi_{\tau}$ are equal to $\pm 1$. The corresponding maximum absolute values of the overall CP asymmetries
(\ref{lg64}) can be written as
\begin{equation}
\begin{array}{rcl}
|\varepsilon^{tot}_{1,\,\mu}|&\simeq &
\dfrac{1}{4\pi}\dfrac{|h^N_{H^u_2 L_3 N_1}||h^N_{H^u_2 L_3 N_2}||A||B|}{2|A|^2+|h^N_{H^u_2 L_3 N_1}|^2}\,\dfrac{M_1}{M_2}\,,\\[4mm]
|\varepsilon^{tot}_{1,\,\tau}|&\simeq &
\dfrac{\biggl(4|h^N_{H^u_2 L_3 N_1}||h^N_{H^u_2 L_3 N_2}||A||B|+3|h^N_{H^u_2 L_3 N_1}|^2 |h^N_{H^u_2 L_3 N_2}|^2\biggr)}
{8\pi\,(2|A|^2+|h^N_{H^u_2 L_3 N_1}|^2)}\dfrac{M_1}{M_2}\,.
\end{array}
\label{lg65}
\end{equation}

In Figs.~6--7 we present the results of our numerical analysis of the decay asymmetries in the E$_6$SSM Model I.
The dependence of the maximum values of $|\varepsilon^{tot}_{1,\,\mu}|$ and $|\varepsilon^{tot}_{1,\,\tau}|$ on $|h^N_{H^u_2 L_3 N_1}|$ and
$|h^N_{H^u_2 L_3 N_2}|$ is studied in Fig.~6. As before we set $(M_2/M_1)=10$, $M_1\simeq 10^6\,\mbox{GeV}$, $v_2\simeq v\simeq 246\,\mbox{GeV}$
and adjust parameters $|A|$ and $|B|$ to reproduce the observed neutrino mass--squared differences.
In Figs.~6a and 6b we plot the maximum value of $|\varepsilon^{tot}_{1,\,\mu}|$ as a function of $|h^N_{H^u_2 L_3 N_1}|$ and $|h^N_{H^u_2 L_3 N_2}|$
while the dependence of the maximum value of $|\varepsilon^{tot}_{1,\,\tau}|$ on these new Yukawa couplings is explored in Figs.~6c and 6d.
From Eq.~(\ref{lg65}) one can see that at very small values of new Yukawa couplings ($|h^N_{H^u_2 L_3 N_1}|,\,|h^N_{H^u_2 L_3 N_2}|\ll |A|$ and
$|B|$) the maximum absolute values of the overall CP asymmetry are proportional to $|h^N_{H^u_2 L_3 N_1}|\cdot|h^N_{H^u_2 L_3 N_2}|$.
At so small values of $|h^N_{H^u_2 L_3 N_1}|$ and $|h^N_{H^u_2 L_3 N_2}|$ the maximum absolute value of the overall CP asymmetry
associated with the decay of $N_1$ into $\tau$--lepton is twice larger than the maximum value of $|\varepsilon^{tot}_{1,\,\mu}|$.
The maximum values of $|\varepsilon^{tot}_{1,\,\mu}|$ and $|\varepsilon^{tot}_{1,\,\tau}|$ rise with increasing of $|h^N_{H^u_2 L_3 N_2}|$
(see Fig.~6a and 6c). When $|h^N_{H^u_2 L_3 N_2}|\gg |A|,\,|B|$ the value of $|\varepsilon^{tot}_{1,\,\tau}|$ tends to be much larger
than $|\varepsilon^{tot}_{1,\,\mu}|$.

At small values of $|h^N_{H^u_2 L_3 N_1}|$ the maximum absolute values of both decay asymmetries also grow with increasing of $|h^N_{H^u_2 L_3 N_1}|$
independently of $|h^N_{H^u_2 L_3 N_2}|$ (see Fig.~6b and 6d). But $|\varepsilon^{tot}_{1,\,\mu}|$ attains its maximum possible value at
$|h^N_{H^u_2 L_3 N_1}|=\sqrt{2}|A|$ whereas $|\varepsilon^{tot}_{1,\,\tau}|$ approach its upper bound at large values of $|h^N_{H^u_2 L_3 N_1}|\gg
|A|,\,|B|$. When $|h^N_{H^u_2 L_3 N_1}|$ is significantly larger than $|A|$ and $|B|$ the the maximum value of $|\varepsilon^t_{1,\,\mu}|$ is
inversely proportional to $|h^N_{H^u_2 L_3 N_1}|$ while $|\varepsilon^{tot}_{1,\,\tau}|$ is almost independent of
$|h^N_{H^u_2 L_3 N_1}|$. In the considered case the theoretical upper bounds on $|\varepsilon^{tot}_{1,\,\mu}|$ and
$|\varepsilon^{tot}_{1,\,\tau}|$ are given by
\begin{equation}
\begin{array}{rcl}
|\varepsilon^{tot}_{1,\,\tau}|&\lesssim &\dfrac{M_1}{8\pi M_2} |h^N_{H^u_2 L_3 N_2}|^2\left[3+\dfrac{4 x}{12+\sqrt{8x+9}}\right]\,,
\qquad x=\dfrac{|B|^2}{|h^N_{H^u_2 L_3 N_2}|^2}\,,\\[3mm]
|\varepsilon^{tot}_{1,\,\mu}|&\lesssim &\dfrac{\sqrt{2} M_1}{16\pi M_2}|h^N_{H^u_2 L_3 N_2}||B|\,.
\end{array}
\label{lg66}
\end{equation}
As before the theoretical restrictions on the absolute values of CP asymmetries (\ref{lg66}) are set by the Yukawa couplings of
the second lightest right--handed neutrino and independent of the Yukawa couplings of the lightest right--handed neutrino.
Because for a fixed ratio $M_1/M_2$ the values of $|A|$ and $|B|\propto \sqrt{M_1}$ the maximum possible value of
$|\varepsilon^{tot}_{1,\,\mu}|$ decreases when $M_1$ becomes smaller while the theoretical upper bound on $|\varepsilon^{tot}_{1,\,\tau}|$
does not change much. As a consequence $|\varepsilon^{tot}_{1,\,\tau}|$ tends to dominate over $|\varepsilon^{tot}_{1,\,\mu}|$
at low masses of the lightest right--handed neutrino $M_1\ll 10^{13}-10^{14}\,\mbox{GeV}$ (see Figs.~6--7).
Since the maximum possible value of $|\varepsilon^{tot}_{1,\,\tau}|$ is determined mainly by $|h^N_{H^u_2 L_3 N_2}|$, which is not
constrained by the neutrino oscillation data, an appreciable CP asymmetry within the E$_6$SSM Model I can be induced even
when $M_1$ is relatively low. Fig.~7 demonstrates that for $M_1\simeq 10^{6}\,\mbox{GeV}$ the decay asymmetry
$|\varepsilon^{tot}_{1,\,\tau}|=10^{-6}-10^{-4}$ can be generated if $|h^N_{H^u_2 L_3 N_2}|$ varies from $0.01$ to $0.1$.

\subsubsection{E$_6$SSM Model II}

Within the E$_6$SSM Model II the lightest right--handed neutrino may decay into the lepto-quarks (squarks) and
down--type squarks (down--type quarks). New decay modes of the lightest right--handed neutrino lead to the set of
extra CP asymmetries $\varepsilon^{i}_{1,\,D_k}$ (\ref{lg38}) which appear in addition to those arising in the E$_6$SSM Model I.
Leptoquarks also give a substantial contribution to $\varepsilon^{k}_{1,\,\ell_x}$, through loop diagrams if the corresponding
Yukawa couplings $g^{N}_{kij}$ are large enough. By construction the exotic quarks and squarks in the E$_6$SSM couple predominantly to the
the quark and lepton superfields of the third generation. Therefore in our analysis we neglect the Yukawa couplings of
the exotic quarks and squarks to the first and second generation particles. Moreover for simplicity we assume that
only the third generation exotic quarks and squarks have appreciable couplings to the bosons and fermions of the third generation
and the Yukawa couplings of $L_4$ and $H^u_{\alpha}$ to the right--handed neutrinos vanish. In this approximation
for the maximum absolute values of the CP asymmetries $|\varepsilon^3_{1,\,\tau}|=|\varepsilon^3_{1,\,\mu}|$ and
$|\varepsilon^3_{1,\,D_3}|$ one obtains
\begin{equation}
|\varepsilon^3_{1,\,\tau}|=|\varepsilon^3_{1,\,\mu}|\simeq\dfrac{3 |B| M_1}{16\pi |A| M_2} |g^{N}_{D_3 d_3 N_1}| |g^{N}_{D_3 d_3 N_2}|\,,
\qquad\qquad |\varepsilon^3_{1,\,D_3}|\simeq\dfrac{3 M_1}{2\pi M_2} |g^{N}_{D_3 d_3 N_2}|^2\,,
\label{lg67}
\end{equation}
where $g^{N}_{D_3 d_3 N_1}\equiv g^{N}_{331}$ and $g^{N}_{D_3 d_3 N_2}=g^{N}_{332}$.
All other decay asymmetries vanish in the considered approximation. As before the maximum absolute values of the
CP asymmetries (\ref{lg67}) tend to zero if $g^{N}_{D_3 d_3 N_1}\to 0$ and $g^{N}_{D_3 d_3 N_2}\to 0$. However in contrast with the scenarios
considered before the absolute values of the CP asymmetries $|\varepsilon^3_{1,\,\mu}|$ and $|\varepsilon^3_{1,\,\tau}|$ do not change
when the lightest right--handed neutrino mass varies while $M_1/M_2$ remains intact. Indeed, according to the Eqs.~(\ref{lg47}) and
(\ref{lg51}) the ratio $|A|/|B|$ is proportional to $\sqrt{M_1/M_2}$. As a result the explicit dependence of the lepton decay asymmetries
on the right--handed neutrino mass scale in Eq.~(\ref{lg67}) is cancelled. The maximum absolute values of the CP asymmetries (\ref{lg67})
are determined by $|g^{N}_{D_3 d_3 N_1}|$ and $|g^{N}_{D_3 d_3 N_2}|$.

The dependence of the maximum values of $|\varepsilon^3_{1,\,\tau}|=|\varepsilon^3_{1,\,\mu}|$ and $|\varepsilon^3_{1,\,D_3}|$ on the
Yukawa couplings $g^{N}_{D_3 d_3 N_1}$ and $g^{N}_{D_3 d_3 N_2}$ is examined in Fig.~8. Once again we fix $(M_2/M_1)=10$,
$v_2\simeq 246\,\mbox{GeV}$ and choose $|A|$ and $|B|$ so that the phenomenologically acceptable pattern of the neutrino mass spectrum
is reproduced. From Eq.~(\ref{lg67}) and Fig.~8 one can see that the decay asymmetries $|\varepsilon^3_{1,\,\tau}|=|\varepsilon^3_{1,\,\mu}|$ and
$|\varepsilon^3_{1,\,D_3}|$ rise monotonically with increasing of $|g^{N}_{D_3 d_3 N_2}|$. The maximum absolute values of the lepton CP asymmetries
also grow when $|g^{N}_{D_3 d_3 N_1}|$ increases. At the same time $|\varepsilon^3_{1,\,D_3}|$ does not depend on $|g^{N}_{D_3 d_3 N_1}|$.
When $|g^{N}_{D_3 d_3 N_2}|\gg |g^{N}_{D_3 d_3 N_1}|$ the decay asymmetry $|\varepsilon^3_{1,\,D_3}|$ tends to be considerably larger than lepton
decay asymmetries. At low energies the induced lepton asymmetry in the exotic quark sector is converted into the ordinary lepton
asymmetries via the decays of leptoquarks into leptons (sleptons) and ordinary quarks (squarks). In the opposite limit
$|g^{N}_{D_3 d_3 N_2}|\ll |g^{N}_{D_3 d_3 N_1}|$ lepton decay asymmetries dominate over $|\varepsilon^3_{1,\,D_3}|$.
If $|g^{N}_{D_3 d_3 N_1}|\sim |g^{N}_{D_3 d_3 N_2}|$ these CP asymmetries are comparable. From Fig.~9 one can see that the appreciable
values of the decay asymmetries $\varepsilon^3_{1,\,\mu}$, $\varepsilon^3_{1,\,\tau}$ and $\varepsilon^3_{1,\,D_3}\sim 10^{-6}-10^{-4}$
can be induced if $|g^{N}_{D_3 d_3 N_1}|,\,|g^{N}_{D_3 d_3 N_2}|\gtrsim 0.01-0.1$.

\subsubsection{Generation of baryon asymmetry}

Although the numerical results for the lepton decay asymmetries look very promissing it is
not clear if an appropriate amount of baryon asymmetry can be generated. In order to
calculate the total lepton and baryon asymmetries produced by thermal leptogenesis in the
considered model the complete system of Boltzmann equations including the ones that describe
the evolution of lepton number densities associated with leptoquarks should be solved.
We plan to derive and analyse a complete set of Boltzmann equations within the E$_6$SSM
in the forthcoming publications.

Nevertheless there is one case which is relatively easy to
analyse. It corresponds to the E$_6$SSM with unbroken $Z^H_2$
symmetry. Indeed, in this case there is only one extra lepton
doublet which interacts with the right--handed neutrinos. The
Yukawa couplings of all other exotic particles to $N_i$ vanish in
the considered limit. Then the complete set of Boltzmann equations
is supplemented by only one extra equation as compared with the
MSSM that describes the evolution of the lepton number density
associated with extra lepton doublet $L_4$. As a consequence all
results obtained in the SM and MSSM for the lepton and baryon
asymmetries can be easily generalised in this case. In particular,
one can estimate the total baryon asymmetry using an approximate
formula (see \cite{review-leptogen})
\begin{equation}
Y_{\Delta B}\sim 10^{-3}\biggl(\sum_{x=1}^4 \varepsilon^3_{1,\,\ell_x}\eta_x\biggr)\,,
\label{lg68}
\end{equation}
where $Y_{\Delta B}$ is a baryon asymmetry relative to the entropy density, i.e.
\begin{equation}
Y_{\Delta B}=\dfrac{n_B-n_{\bar{B}}}{s}\biggl|_0=(8.75\pm
0.23)\times 10^{-11}\,. \label{lg69}
\end{equation}
In Eq.~(\ref{lg68}) $\eta_x$ is an efficiency factor. A thermal population of
$N_1$ and $\tilde{N}_1$ decaying completely out of equilibrium without washout effects
would lead to $\eta_{x}=1$. However inverse decays and other washout processes  reduce the
induced asymmetries by factor $\eta_x$ where $\eta_x$ varies from 0 to 1\footnote{$\eta_x=0$
is the limit of $N_1$ interactions in perfect equilibrium so that no asymmetry is created.}.

\begin{table}[ht]
  \centering
  \begin{tabular}{|c|c|c|c|c|c|c|c|}
    \hline
& $|h^{N}_{H^u_3 L_4 N_1}|$   & $|h^{N}_{H^u_3 L_4 N_2}|$ &
$|\varepsilon^3_{1,\,L_4}|_{max}$ &
$|\varepsilon^3_{1,\,\tau}|_{max}$ &
$\eta_{L_4}$ & $\eta_{\mu}=\eta_{\tau}$ & $|Y_{\Delta B}|_{max}$ \\
& & & & $=|\varepsilon^3_{1,\,\mu}|_{max}$ & & & \\
\hline (A) & 0.1 & 1.0 & 0.0119 & $5.42\cdot 10^{-11}$ &
$3.26\cdot 10^{-9}$ & 0.077 & $3.9\cdot 10^{-14}$\\
\hline (B) & $10^{-3}$ & 1.0 & 0.0119 & $5.42\cdot 10^{-9}$ &
$3.26\cdot 10^{-5}$ & 0.077 & $3.9\cdot 10^{-10}$\\
\hline (C) & $10^{-5}$ & 1.0 & 0.00127 & $5.75\cdot 10^{-8}$ &
0.326 & 0.077 & $4.1\cdot 10^{-7}$\\
\hline (D) & 0.1 & 0.01 & $1.2\cdot 10^{-6}$ & $5.42\cdot
10^{-13}$ &
$3.26\cdot 10^{-9}$ & 0.077 & $3.9\cdot 10^{-18}$\\
\hline (E) & $10^{-3}$ & 0.01 & $1.2\cdot 10^{-6}$ & $5.42\cdot
10^{-11}$ &
$3.26\cdot 10^{-5}$ & 0.077 & $3.9\cdot 10^{-14}$\\
\hline (F) & $10^{-5}$ & 0.01 & $1.27\cdot 10^{-7}$ & $5.75\cdot
10^{-10}$ &
0.326 & 0.077 & $4.1\cdot 10^{-11}$\\
\hline
\end{tabular}
  \caption{\it\small Lepton decay asymmetries, efficiency factors and baryon
asymmetry estimated for $M_1=10^6\,\mbox{GeV}$, $M_2=10 M_1$ and
different values of $|h^{N}_{H^u_3 L_4 N_1}|$ and
$|h^{N}_{H^u_3L_4 N_2}|$.}
  \label{asym}
\end{table}

As in the SM and MSSM it is convenient to introduce a set of
dimensionful parameters
\begin{equation}
\widetilde{m}_{1,x}=h^{N}_{H^u_3 L_x N_1} h^{N*}_{H^u_3 L_x N_1}\dfrac{v_2^2}{M_1}\,,
\qquad\qquad \widetilde{m}_{1}=\sum_{x=1}^4 \widetilde{m}_{1,x}\,,
\label{lg70}
\end{equation}
\begin{equation}
m_{*}=8\pi\dfrac{v_2^2}{M_1^2}H\biggl|_{T=M_1}\,,\qquad\qquad\qquad
H(T=M_1)=1.66 g_{*}^{1/2}\dfrac{T^2}{M_{Pl}}\biggl|_{T=M_1}\,,
\label{lg71}
\end{equation}
where $H$ is a Hubble expansion rate and $g_{*}=n_b+\dfrac{7}{8}\,n_f$ is 
a number of relativistic degrees of freedom in the thermal bath. Within 
the SM $g_{*}=106.75$ and $m_{*}=1.08\cdot 10^{-3}\,\mbox{eV}$ while in
the E$_6$SSM $g_{*}=356.25$ and $m_{*}=1.97\cdot
10^{-3}\,\mbox{eV}$. Here we concentrate on the so--called strong
washout scenario when $\widetilde{m}_{1,x}>m_{*}$. In this case
the efficiency factor $\eta_x$ for flavor $\ell_x$ can be
estimated as (see \cite{review-leptogen})
\begin{equation}
\eta_x\simeq \dfrac{m_{*}}{\widetilde{m}_{1,x}}\,. \label{lg72}
\end{equation}

The results of our numerical studies are summarised in Table 1. We
compute the maximal absolute values of the lepton decay
asymmetries and estimate the efficiency factors as well as induced
baryon asymmetry for different values of $|h^{N}_{H^u_3 L_4 N_1}|$
and $|h^{N}_{H^u_3 L_4 N_2}|$ within the see--saw models with
constrained sequential dominance. As before we set
$M_1=10^6\,\mbox{GeV}$, $M_2=10 M_1$ and calculate the Yukawa
couplings of the Higgs field $H_u$ to ordinary lepton doublets and
right--handed neutrinos assuming CSD. We restrict our
consideration to the part of the parameter space where
$|\varepsilon^3_{1,\,L_4}|$ can be relatively large, i.e. $\gtrsim
10^{-6}$. This corresponds to the $|h^{N}_{H^u_3 L_4 N_2}|\gtrsim
0.01$ (see Fig.~5b). The results presented in Table 1 indicate
that in the scenarios (B), (C) and (F) a substantial amount of
baryon asymmetry may be generated. At the same time one can see
that in some cases when the maximal value of
$|\varepsilon^3_{1,\,L_4}|$ is relatively large the efficiency
factor is so small that almost all lepton asymmetry is erased.
Therefore the generation of relatively large lepton decay
asymmetries does not guarantee the successful baryogenesis in the
E$_6$SSM. This is a necessary condition but not a sufficient one.
We will perform a detailed analysis of the generation of
baryon and lepton asymmetries within the E$_6$SSM in the near
future.

\section{Conclusions}

In~ this~ paper~ we~ have~ discussed~ the~ mechanism~ of~ generation~ of~ lepton~ asymmetry~ within~ the~ Exceptional~
Supersymmetric~ Standard~ Model.~ The~ E$_6$SSM~ is~ based~ on~\\ the~ $SU(3)_C\times SU(2)_W\times U(1)_Y\times U(1)_N$
gauge group which is a subgroup of $E_6$. The particle content of the Exceptional SUSY model includes three complete fundamental
representations of $E_6$ as well as the doublet $L_4$ and anti--doublet $\overline{L}_4$ from an extra $27'$ and $\overline{27'}$.
Thus the E$_6$SSM involves exotic matter beyond the MSSM. In particular, it predicts the existence of
three generations of exotic quarks $D_i$ and $\overline{D}_i$, and two generations of inert Higgs fields $H^d_{\alpha}$ and
$H^u_{\alpha}$ that do not carry any lepton or baryon number as well as a vector-like fourth
lepton doublet $L_4$ and $\overline{L}_4$ that carry a lepton number $L=\pm 1$.
In the phenomenologically
acceptable $E_6$ inspired models the extra exotic quarks can be either diquarks (E$_6$SSM Model I) or
leptoquarks (E$_6$SSM Model II).
In order to suppress the couplings of extra inert Higgs fields to ordinary quarks and leptons that lead to
the unacceptably large non--diagonal flavour transitions we imposed an approximate $Z^H_2$ symmetry.

In the E$_6$SSM right--handed neutrinos and their superpartners do not participate in the gauge interactions and therefore
can be significantly heavier than other particles from the same $27$--plet. The right--handed neutrino mass scale is not
fixed in the considered model but it is expected to be much lower than the Grand Unification scale. The three known doublet
neutrinos $\nu_e$, $\nu_{\mu}$ and $\nu_{\tau}$ acquire small Majorana masses via the seesaw mechanism in this case.
Because right--handed neutrinos are allowed to have large masses in the considered model, they may decay into
the final states with lepton number $L=\pm 1$, thereby creating a lepton asymmetry in the early Universe. The dynamically
induced lepton asymmetry subsequently gets converted into the observed baryon asymmetry through the electroweak phase
transition. The process of the lepton asymmetry generation is controlled by the flavour
dependent CP (decay) asymmetries.
These decay asymmetries originate from the interference of the tree--level and one--loop amplitudes of the lightest
right--handed neutrino decays. We have calculated flavour CP asymmetries within the Exceptional SUSY model and
analysed their dependence on the Yukawa couplings of new exotic particles.

The new exotic particles predicted by the E$_6$SSM contribute to the ordinary CP asymmetries induced by the
decays of the lightest right--handed neutrino (or sneutrino) into the final states containing leptons and sleptons
through loop diagrams. The new particles and interactions also result in new channels of the decays of the lightest
right--handed neutrino and its superpartner which give rise to a set of extra decay asymmetries associated with
new decay modes. When $Z^H_2$ symmetry is unbroken the only exotic particles that contribute to the generation
of lepton asymmetry are the fermion and scalar components of the vector-like lepton doublet superfield $L_4$.
In this case the analytic expressions for the flavour CP asymmetries obtained in the MSSM can be easily generalised
to include $L_4$ since this field may be considered as a fourth generation lepton doublet.
As a result in the E$_6$SSM with unbroken $Z^H_2$ symmetry there are four independent CP asymmetries.
These decay asymmetries take on non--zero values only if CP invariance is broken either in the lepton or
$L_4$ sectors, i.e. some of the Yukawa couplings of either leptons or $L_4$ are complex. Our numerical analysis
reveals that the absolute values of the ordinary lepton CP asymmetries diminish when the mass of the lightest right--handed
neutrino decreases. This happens because the values of these decay asymmetries are set by the Yukawa couplings
of leptons to the Higgs doublet $H_u$ and right--handed neutrinos which also determine the Majorana masses of the
left--handed neutrino. Then the pattern of the neutrino mass--squared differences measured in the neutrino oscillation
experiments imply that the lower masses of the right--handed neutrinos require the smaller values of the corresponding
Yukawa couplings. As a consequence at low right--handed neutrino mass scales ($M_1\ll 10^{9}\,\mbox{GeV}$)
the ordinary lepton decay asymmetries become extremely small.

The CP asymmetries also depend rather strongly on the Yukawa couplings of $L_4$. In contrast with
the couplings of the ordinary lepton fields to the Higgs doublet $H_u$ and right--handed neutrinos the similar Yukawa
couplings of $L_4$ are not constrained by the neutrino oscillation data because they are not related with the Majorana
masses of the left--handed neutrino. Therefore these couplings can vary within a very wide range at the lightest
right--handed neutrino mass scale. If the Yukawa couplings of the fourth lepton doublet superfield $L_4$ are large and complex
they can induce a substantial CP asymmetry associated with the decay $N_1\to L_4+H_{u}$ ($|\varepsilon_{1,\,L'}|\sim 10^{-4}$)
even for relatively low $M_1$ (for example, $M_1\simeq 10^6\,\mbox{GeV}$). When $M_1\ll 10^{13}-10^{14}\,\mbox{GeV}$
this CP asymmetry tends to dominate over the ordinary lepton decay asymmetries. At low energies the lepton asymmetry
generated in the $L_4$ sector gets converted into the ordinary lepton asymmetries via the decays of $L_4$.
We have derived theoretical restrictions on the absolute values of the CP asymmetries within the see--saw models with
constrained sequential dominance. The corresponding theoretical upper bounds are determined by the Yukawa couplings of
the second lightest right--handed neutrino and do not depend on the Yukawa couplings of the lightest
right--handed neutrino.

In the E$_6$SSM Model I the interactions of inert Higgs superfields $H^u_{\alpha}$ with leptons are allowed. These
interactions give rise to eight extra decay asymmetries which are not present in the E$_6$SSM with unbroken $Z^H_2$ symmetry.
We defined four total flavour CP asymmetries which correspond to three different lepton flavours ($e$, $\mu$, $\tau$)
and fourth lepton doublet $L_4$. It is natural to assume that inert Higgs fields
$H^u_{\alpha}$ couple predominantly to the third generation fermions whereas the couplings of $H^u_{\alpha}$ to the
quarks and leptons of the first two generation are negligibly small. Note that neither neutrino nor collider experiments
set any limit on the Yukawa couplings of the inert--Higgs fields $H^u_{\alpha}$ to $\tau$--lepton and
right--handed neutrinos so that these couplings can be of the order of $0.1$\,. In the considered approximation the inert--Higgs
supermultiplets $H^u_{\alpha}$ can give a considerable contribution to the total decay asymmetry associated with $\tau$--lepton
only. As before the absolute values of muon and electron CP asymmetries reduce when the mass of the lightest right--handed
neutrino decreases. At low right--handed neutrino mass scales these asymmetries are negligibly small.
At the same time if the Yukawa couplings of $H^u_{\alpha}$ to the $\tau$--lepton and right--handed neutrinos
are large and complex they can induce a substantial total decay asymmetry associated with $\tau$--lepton which
tends to dominate over the electron and muon ones when $M_1\ll 10^{13}-10^{14}\,\mbox{GeV}$. We have established the
theoretical upper bounds on the absolute values of the total flavour CP asymmetries in the framework of constrained
sequential dominance. Again the maximal absolute values of the total decay asymmetries are set by the Yukawa couplings of
the second lightest right--handed neutrino. Our analysis demonstrates that an appreciable $\tau$--lepton CP asymmetry
$|\varepsilon^{tot}_{1,\,\tau}|\simeq 10^{-6}-10^{-4}$ can be generated even if $M_1\simeq 10^6\,\mbox{GeV}$.

Many new decay channels of the lightest right--handed neutrino appear within the E$_6$SSM Model II. In this model
the exotic quarks and squarks carry lepton and baryon number simultaneously, i.e. they are leptoquarks.
As a result the decays of the lightest right--handed neutrino into leptoquarks (squarks) and down--type squarks
(quarks) are allowed. New decay modes of $N_1$ lead to the nine extra CP asymmetries in addition to those
arising in the E$_6$SSM Model I. Here we defined three extra total CP asymmetries which correspond to three
generations of leptoquarks. The values of extra CP asymmetries are determined by the Yukawa couplings of
leptoquarks which are not constrained by either neutrino or collider experiments. If these Yukawa couplings are
large and complex they give rise to the appreciable CP asymmetries associated with the exotic quarks which
are converted into the ordinary lepton asymmetries via the decays of leptoquarks into leptons (sleptons)
at low energies. In the considered case the exotic quarks and squarks can
also give a large contribution to the lepton CP asymmetries through loop diagrams so that a substantial values
of $\varepsilon^k_{1,\,\ell_x}$ are induced even for the relatively low right--handed neutrino mass scales.
In general the considerable values of CP asymmetries $\sim 10^{-6}-10^{-4}$ can be induced
independently of the right--handed neutrino mass scale if new Yukawa couplings of the
right--handed neutrino vary from $0.01$ to $0.1$.

Thus the results presented in this paper show that substantial asymmetries can be
generated even for very low right--handed neutrino masses of order $10^6\,\mbox{GeV}$.
We have also briefly considered the efficiency factors relevant for the
case of the E$_6$SSM with unbroken $Z_2^H$ symmetry, and shown that
acceptably large baryon asymmetries can result. These results suggest that in the
E$_6$SSM successful thermal leptogenesis can be achieved without encountering
problems with gravitinos.


\section*{Acknowledgements}
\vspace{0mm} We would like to thank S.~Antusch,
C.~D.~Froggatt, G.~G.~Ross, S.~Sarkar, J.~March-Russell, L.~B.~Okun,
P.~Soler and D.~Sutherland for fruitful discussions. RN acknowledge support from the SHEFC grant HR03020 SUPA 36878.
SFK acknowledges partial support from the following grants:
PPARC Rolling Grant PPA/G/S/2003/00096;
EU Network MRTN-CT-2004-503369;
NATO grant PST.CLG.980066;
EU ILIAS RII3-CT-2004-506222.

\newpage

\section*{Appendix: CP asymmetries in the effective field\\ theory approach}

\setcounter{equation}{0}
\def\theequation{A.\arabic{equation}}

In the presence of a large mass gap between the lightest and second lightest right--handed neutrinos one can use the
effective field theory approach for the calculation of CP asymmetries. Assuming that $M_2,\, M_3\gg M_1$
one can integrate out two heavy right--handed neutrinos so that at the energies below $M_2$
but above $M_1$ the interactions between superfields are described by the effective superpotential:
\begin{equation}
\begin{array}{c}
W_{\rm eff}\simeq h^{N}_{kx1}(H^u_{k} L_x) N_1^c+ g^N_{ki1}(D_k d^c_i) N_1^c+\dfrac{1}{2}\,\Xi^{LL}_{kxiy}(H^u_{k} L_x) (H^u_{i} L_y)\\[2mm]
+\Xi^{LD}_{kxmp}(H^u_{k} L_x) (D_m d^c_p)+\dfrac{1}{2}\,\Xi^{DD}_{kimp} (D_k d^c_i)(D_m d^c_p)+...\,,
\end{array}
\label{lg411}
\end{equation}
where $i$, $k$, $m$, $p$ are family indexes that run from 1 to 3 while $x$ and $y$ vary from 1 to 4 and
\begin{equation}
\begin{array}{c}
\Xi^{LL}_{kxiy}=\Xi^{LL}_{iykx}=-\sum_{j=2,3}\dfrac{h^{N}_{kxj} h^{N}_{iyj}}{M_j}\,,\qquad\qquad
\Xi^{LD}_{kxmp}=-\sum_{j=2,3}\dfrac{h^{N}_{kxj}g^{N}_{mpj}}{M_j}\,,\\[2mm]
\Xi^{DD}_{kimp}=\Xi^{DD}_{mpki}=-\sum_{j=2,3}\dfrac{g^{N}_{kij}g^{N}_{mpj}}{M_j}\,.
\end{array}
\label{lg412}
\end{equation}
The last three terms in the superpotential (\ref{lg411}) are the lowest dimensional effective operators
induced by the heavy right--handed neutrinos. In the effective field theory approach these operators
give rise to the non--zero values of the CP asymmetries which originate from the interference of the
tree--level and one--loop amplitudes of the lightest right--handed neutrino decays.
The associated one--loop diagrams involve two vertices one of which represents non--renormalisable lepton number violating
interactions induced by the heavy right--handed neutrinos. Calculating one--loop diagrams, we find
\begin{equation}
\begin{array}{rcl}
\varepsilon^{k}_{1,\,\ell_x}&\simeq &\dfrac{M_1}{8\pi A_1}
\mbox{Im}\biggl[2\sum_{i,\,y} h^{N*}_{kx1}\,\Xi^{LL}_{kxiy}\,h^{N*}_{iy1}+
\sum_{m,\,p} h^{N*}_{kx1}\,\Xi^{LL}_{ixky}\, h^{N*}_{iy1}\\[2mm]
&+& 3\sum_{m,\,p} h^{N*}_{kx1}\,\Xi^{LD}_{kxmp}\, g^{N*}_{mp1}\biggr]\,,\\[2mm]
\varepsilon^{i}_{1,\,D_k}&\simeq &\dfrac{M_1}{8\pi A_0}
\mbox{Im}\biggl[3\sum_{m,\,p} g^{N*}_{ki1}\,\Xi^{DD}_{kimp}\, g^{N*}_{mp1}+
\sum_{m,\,p} g^{N*}_{ki1}\,\Xi^{DD}_{mikp}\, g^{N*}_{mp1}\\
&+&2\sum_{m,\,p} g^{N*}_{ki1}\,\Xi^{LD}_{myki}\, h^{N*}_{my1}\biggr]\,,
\end{array}
\label{lg413}
\end{equation}
In the case of the E$_6$SSM $\mbox{version I}$, $g^{N}_{mpj}$, $\Xi^{DD}_{kimp}$ and $\Xi^{LD}_{kxim}$ vanish
and the expressions for the decay asymmetries (\ref{lg413}) are simplified drastically. In particular,
$\varepsilon^{i}_{1,\,D_k}$ tend to zero because in the considered case the exotic quarks are diquarks and therefore
the baryon number conservation forbids the decay of the lightest right--handed neutrino into exotic quarks (or squarks).
The Eqs.~(\ref{lg413}) can be simplified even further if one sets  $h^N_{1xj}=h^N_{2xj}=0$. Then
the results for the CP asymmetries (\ref{lg21}) derived in the exact $Z_2^{H}$ symmetry limit are reproduced.
Finally, the Eqs.~(\ref{lg413}) can be obtained directly from Eqs.~(\ref{lg30}) and (\ref{lg39})
by setting $(M_1/M_j)\to 0$. One can easily check that in the considered limit the analytic expressions for the decay
asymmetries (\ref{lg30}) and (\ref{lg39}) coincide with the results obtained in the effective field theory
approximation.

\renewcommand{\baselinestretch}{1.00}

\begin{figure}
\includegraphics[width=65mm,clip=true]{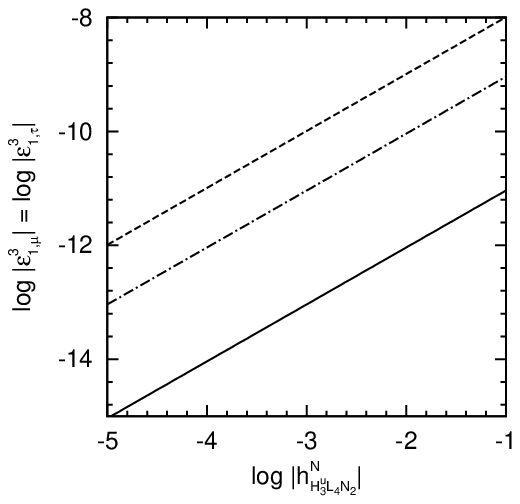} \hspace*{5mm}
\includegraphics[width=65mm,clip=true]{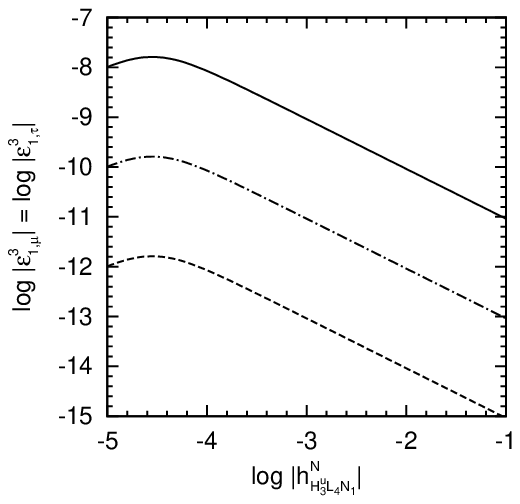}
\hspace*{37mm}{\bf (a)}\hspace*{88mm}{\bf (b) }\\[3mm]
\includegraphics[width=65mm,clip=true]{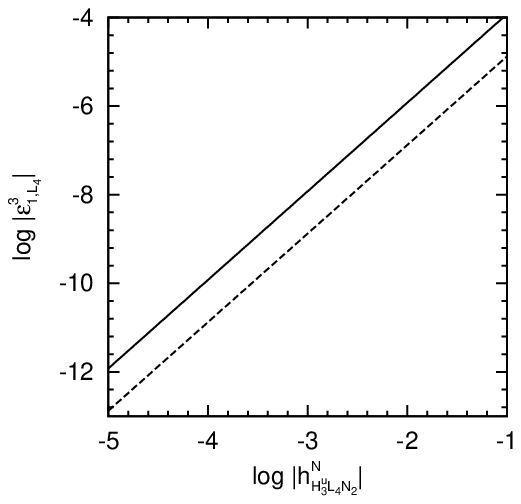} \hspace*{5mm}
\includegraphics[width=65mm,clip=true]{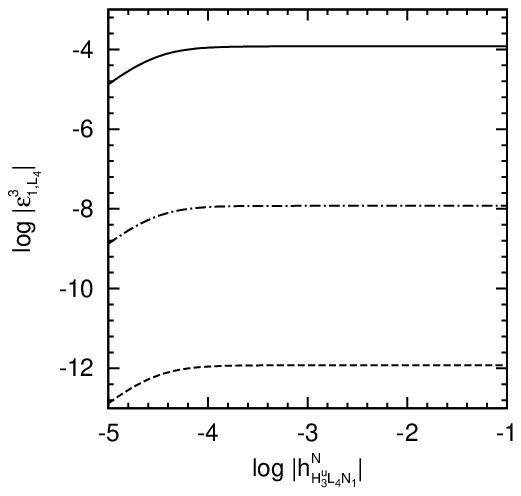}
\hspace*{37mm} {\bf (c)}\hspace*{88mm}{\bf (d) }\\[3mm]
\caption{Maximal absolute values of {\it (a)}--{\it (b)} $|\varepsilon^3_{1,\,\mu}|=|\varepsilon^3_{1,\,\tau}|$
and {\it (c)}--{\it (d)} $|\varepsilon^3_{1,\,L_4}|$ in the E$_6$SSM with unbroken $Z^2_H$ symmetry
versus $|h^N_{H^u_3 L_4 N_1}|$ and $|h^N_{H^u_3 L_4 N_2}|$ for $M_1=10^{6}\,\mbox{GeV}$, $M_2=10\cdot M_1$.
The solid, dash--dotted and dashed lines in figures {\it (a)} and {\it (c)} represent the maximal
absolute values of the decay asymmetries for $|h^N_{H^u_3 L_4 N_1}|=0.1,\,10^{-3}$ and $10^{-5}$ while
solid, dash--dotted and dashed lines in figures {\it (b)} and {\it (d)} correspond to
$|h^N_{H^u_3 L_4 N_2}|=0.1,\,10^{-3}$ and $10^{-5}$ respectively.}
\end{figure}

\begin{figure}
\begin{center}
\includegraphics[width=75mm,clip=true]{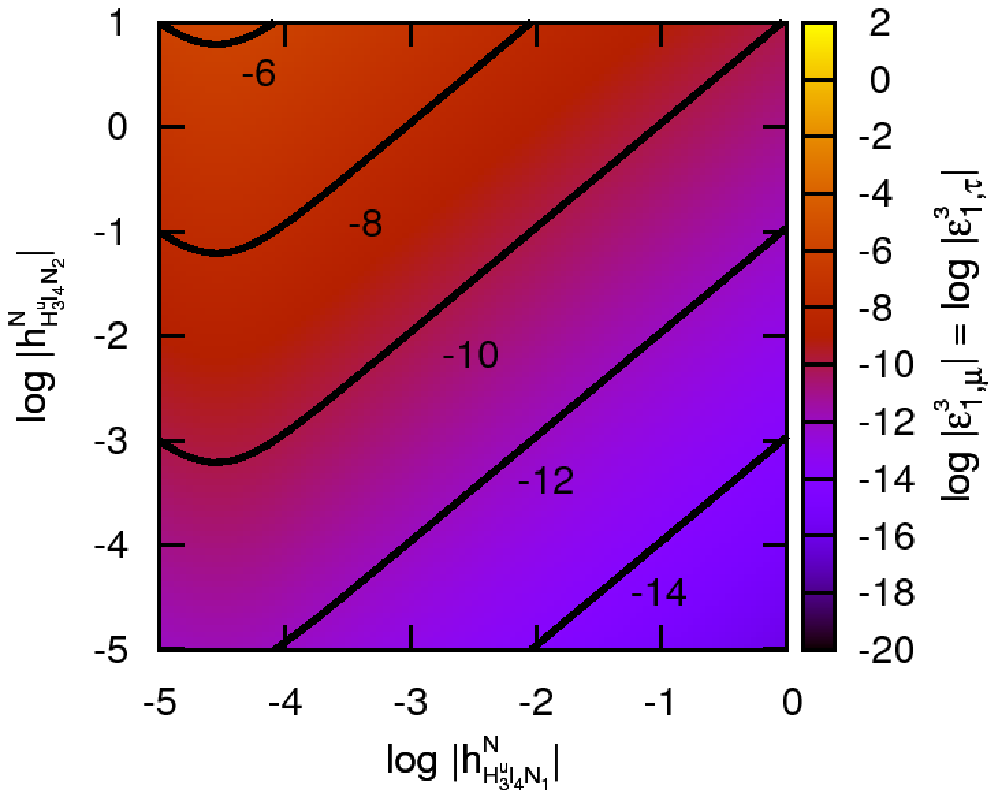} \hspace{5mm}
\includegraphics[width=75mm,clip=true]{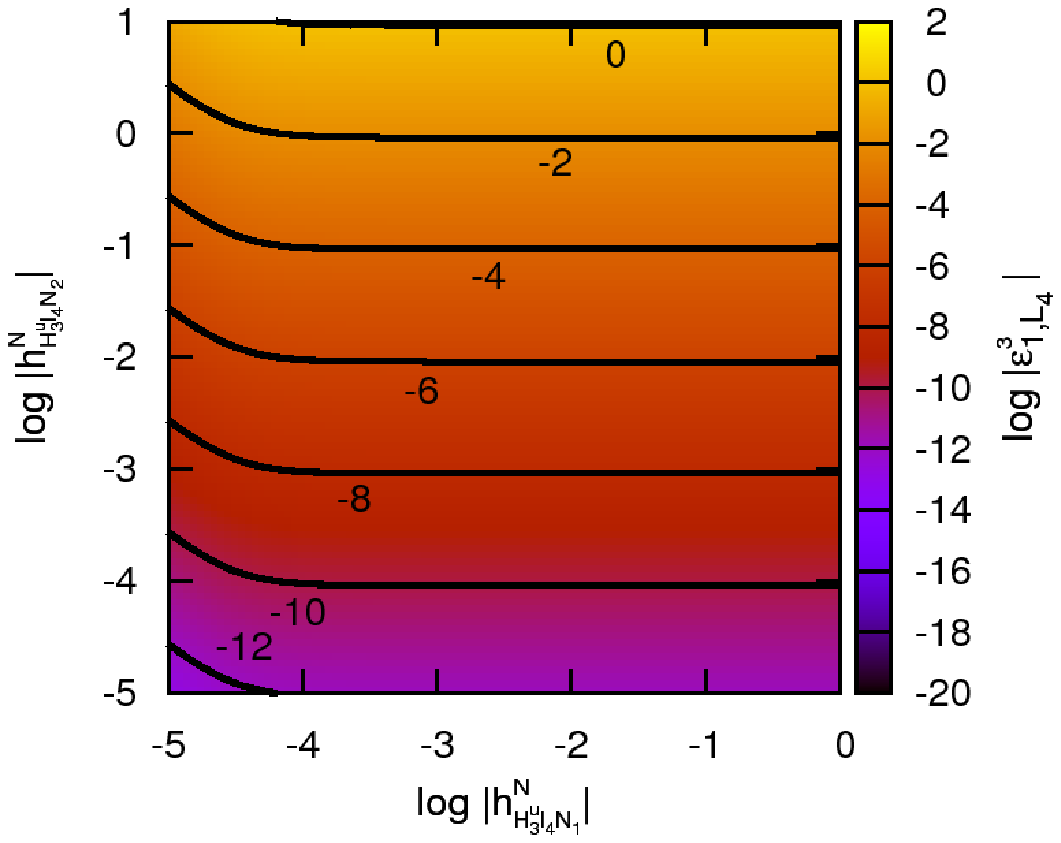}
{\bf (a)}\hspace*{80mm}{\bf (b) }\\[3mm]
\includegraphics[width=75mm,clip=true]{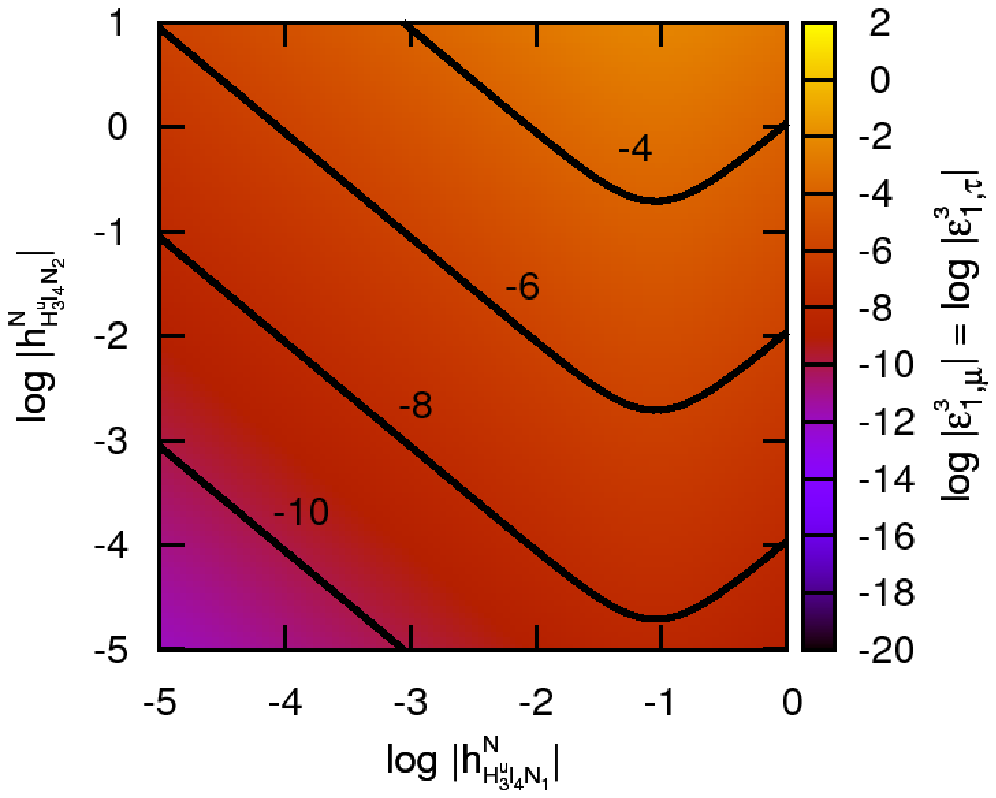} \hspace{5mm}
\includegraphics[width=75mm,clip=true]{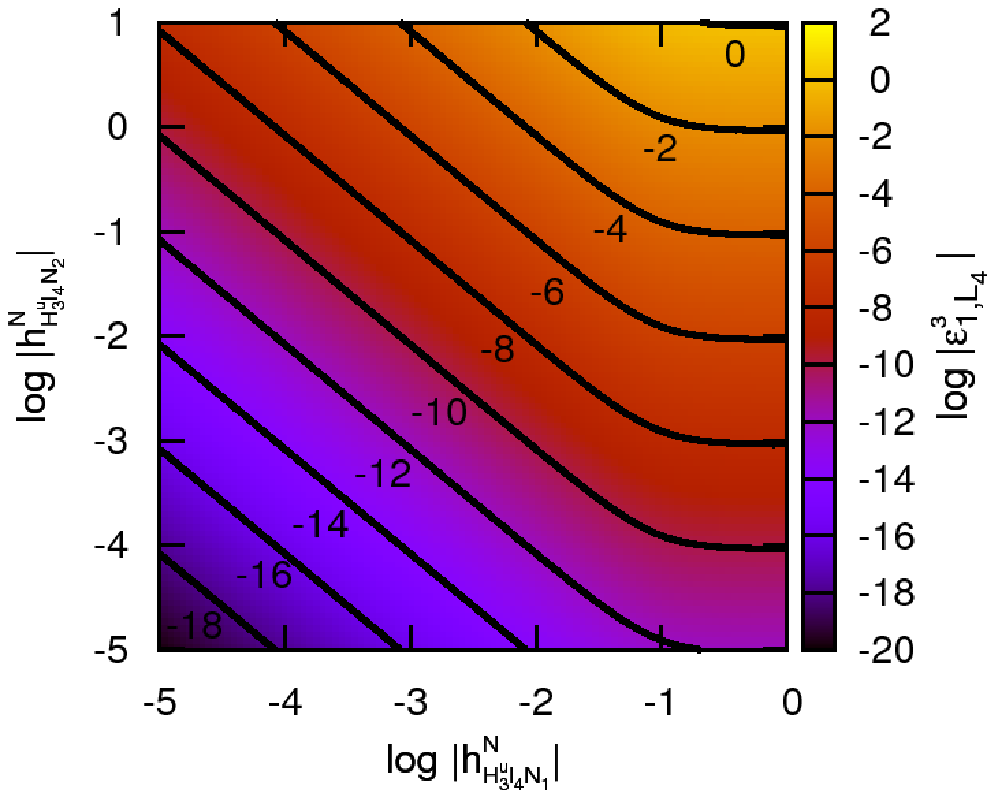}
{\bf (c)}\hspace*{80mm}{\bf (d) }\\[3mm]
\caption{ Logarithm (base 10) of the maximal values of $|\varepsilon^3_{1,\,\mu}|=|\varepsilon^3_{1,\,\tau}|$ ({\it a, c}) and
$|\varepsilon^3_{1,\,L_4}|$ ({\it b, d}) in the E$_6$SSM with unbroken $Z^2_H$ symmetry versus $\log |h^N_{H^u_3 L_4 N_1}|$
and $\log |h^N_{H^u_3 L_4 N_2}|$ for $M_1=10^{6}\,\mbox{GeV}$ ({\it a, b}), $M_1=10^{13}\,\mbox{GeV}$ ({\it c, d}), and $M_2=10\, M_1$.
The solid contour lines show steps of $2$ in the logarithm (base 10) of the asymmetries.}
\end{center}
\end{figure}

\begin{figure}
\includegraphics[width=65mm,clip=true]{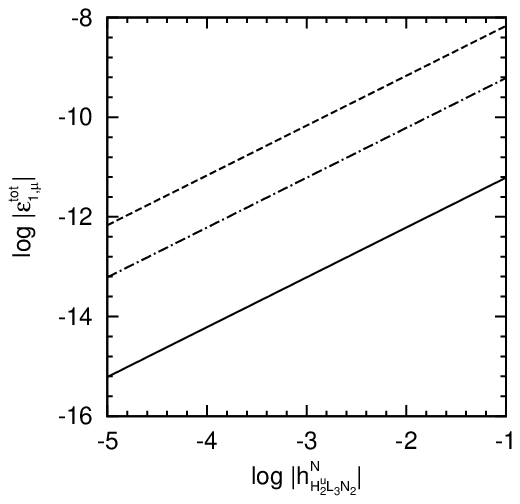} \hspace*{5mm}
\includegraphics[width=65mm,clip=true]{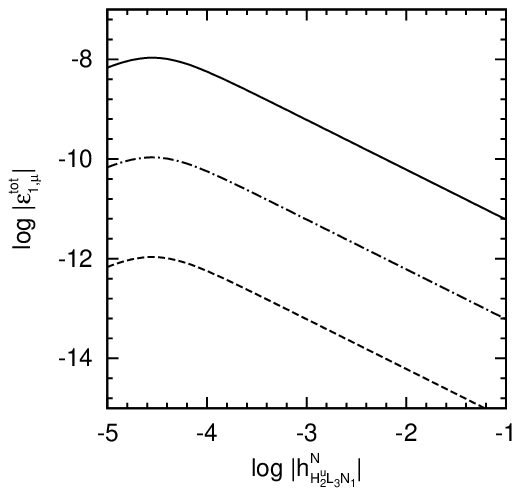}
\hspace*{37mm}{\bf (a)}\hspace*{88mm}{\bf (b) }\\[3mm]
\includegraphics[width=65mm,clip=true]{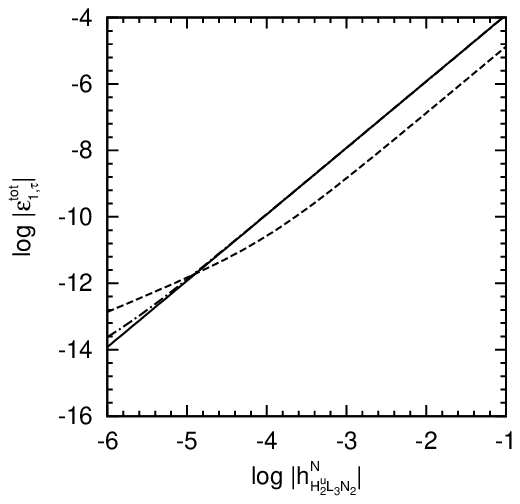} \hspace*{5mm}
\includegraphics[width=65mm,clip=true]{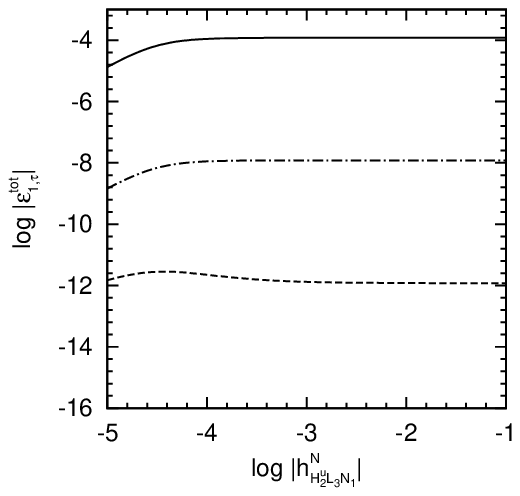}
\hspace*{37mm}{\bf (c)}\hspace*{88mm}{\bf (d) }\\[3mm]
\caption{Maximal absolute values of {\it (a)}--{\it (b)} $|\varepsilon^{tot}_{1,\,\mu}|$
and {\it (c)}--{\it (d)} $|\varepsilon^{tot}_{1,\,\tau}|$ in the E$_6$SSM Model I
versus $|h^N_{H^u_2 L_3 N_2}|$ and $|h^N_{H^u_2 L_3 N_1}|$ for $M_1=10^{6}\,\mbox{GeV}$ and $M_2=10\cdot M_1$.
All couplings $|h^N_{H^u_k L_4 N_j}|$ are set to zero. The solid, dash--dotted and dashed lines
in figures {\it (a)} and {\it (c)} represent the maximal absolute values of the decay
asymmetries for $|h^N_{H^u_2 L_3 N_1}|=0.1,\,10^{-3}$ and $10^{-5}$ while solid, dash--dotted and
dashed lines in figures {\it (b)} and {\it (d)} correspond to $|h^N_{H^u_2 L_3 N_2}|=0.1,\,10^{-3}$
and $10^{-5}$ respectively.}
\end{figure}

\begin{figure}
\begin{center}
\includegraphics[width=75mm,clip=true]{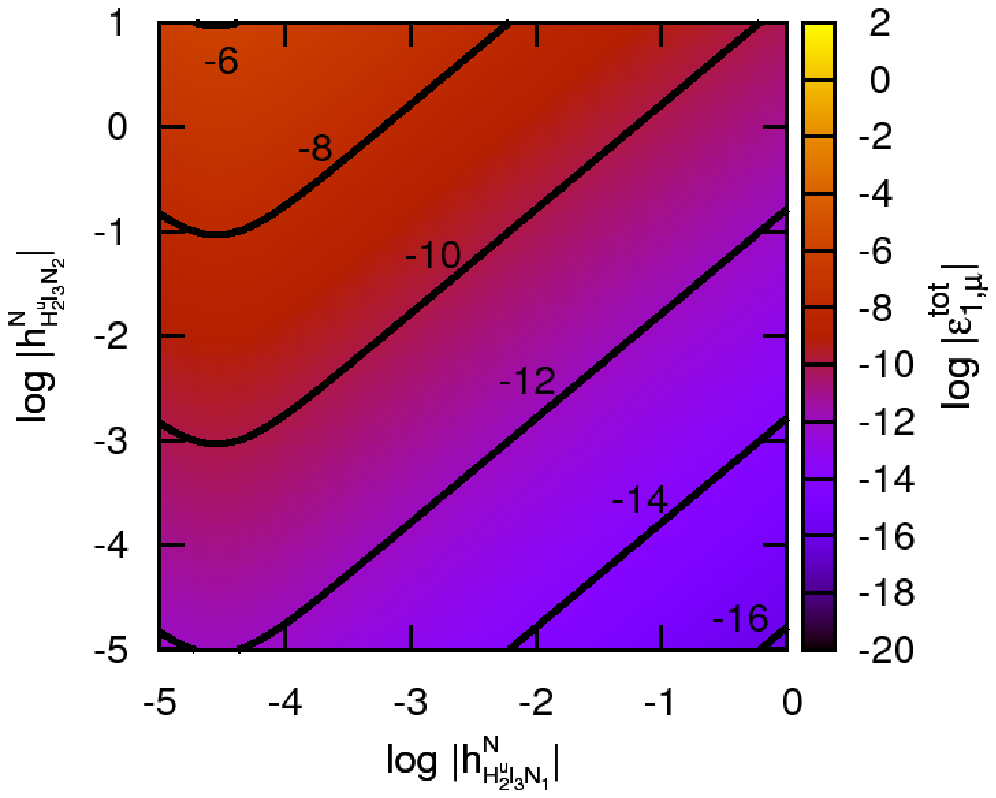} \hspace{5mm}
\includegraphics[width=75mm,clip=true]{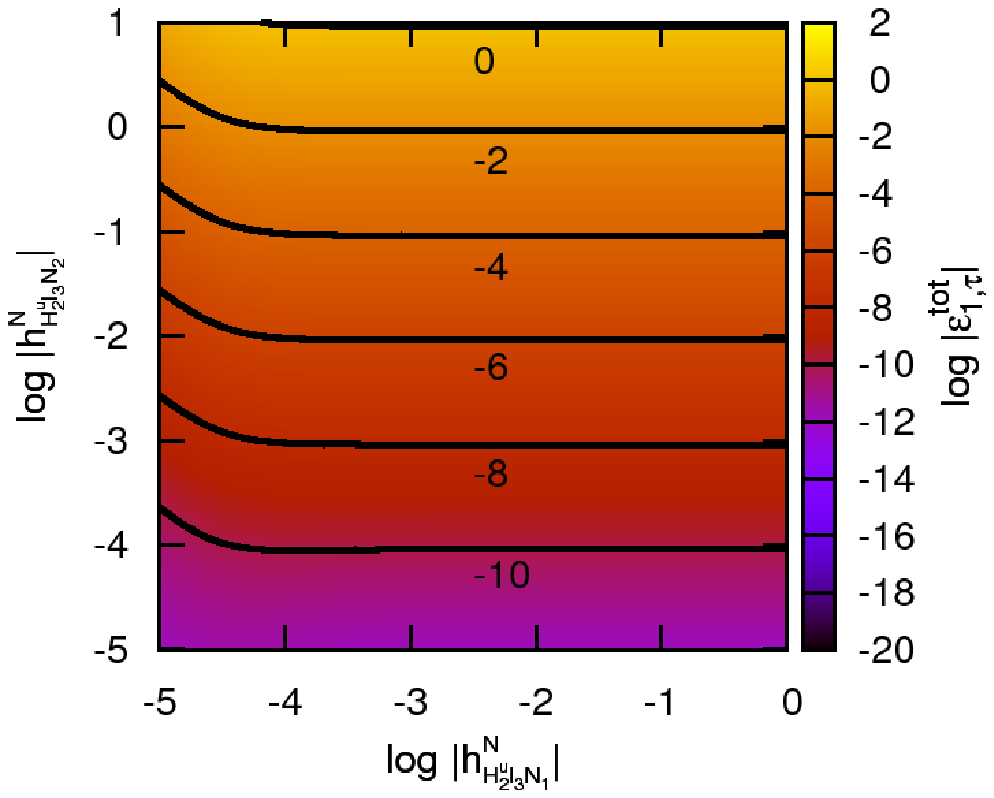}
{\bf (a)}\hspace*{80mm}{\bf (b) }\\[3mm]
\includegraphics[width=75mm,clip=true]{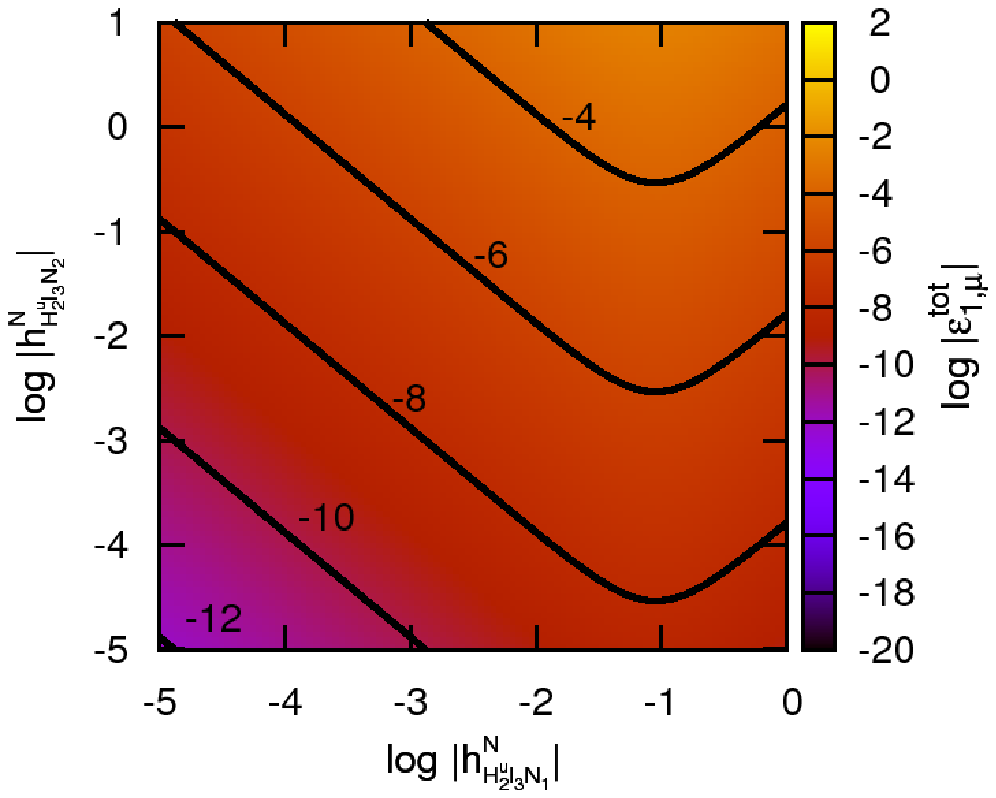} \hspace{5mm}
\includegraphics[width=75mm,clip=true]{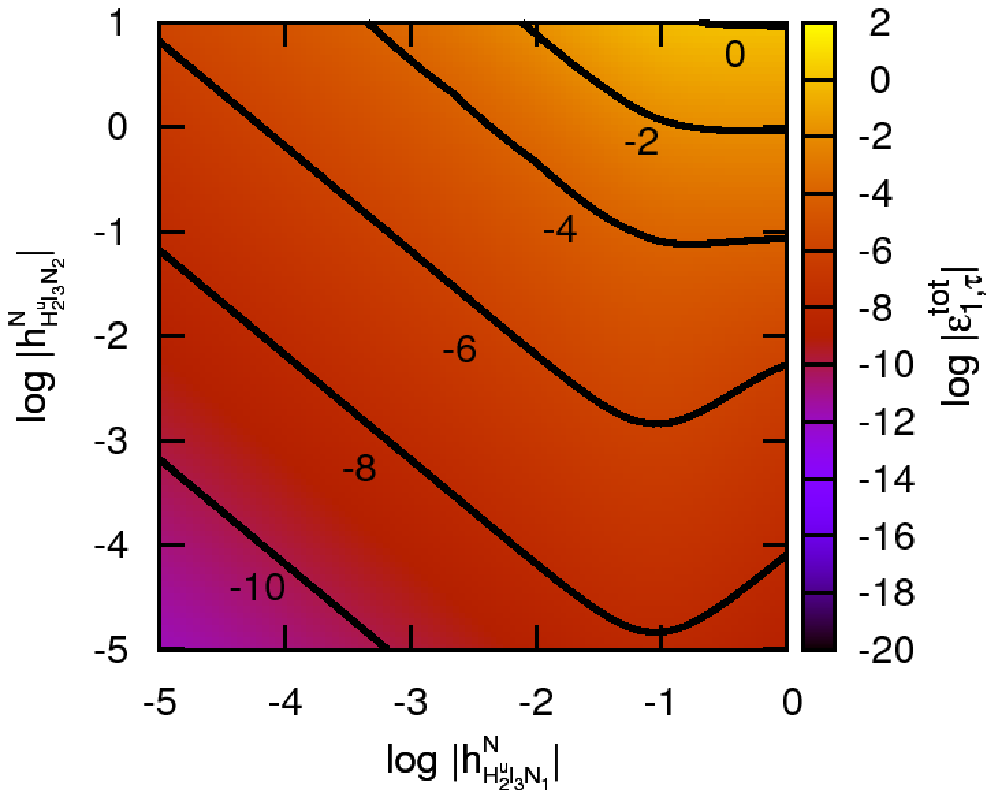}
{\bf (c)}\hspace*{80mm}{\bf (d) }\\[3mm]
\caption{ Logarithm (base 10) of the maximal values of $|\varepsilon^{tot}_{1,\,\mu}|$ ({\it a, c}) and
$|\varepsilon^{tot}_{1,\,\tau}|$ ({\it b, d}) in the E$_6$SSM Model I versus $\log |h^N_{H^u_2 L_3 N_2}|$
and $\log |h^N_{H^u_2 L_3 N_1}|$ for $M_1=10^{6}\,\mbox{GeV}$ ({\it a, b}), $M_1=10^{13}\,\mbox{GeV}$
({\it c, d}), and $M_2=10\, M_1$. All couplings $|h^N_{H^u_k L_4 N_j}|$ are set to zero.
The solid contour lines show steps of $2$ in the logarithm (base 10) of the asymmetries.}
\end{center}
\end{figure}

\newpage
\begin{figure}
\includegraphics[width=65mm,clip=true]{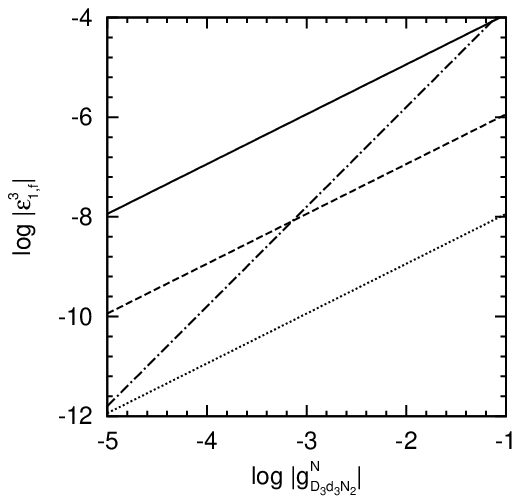} \hspace*{5mm}
\includegraphics[width=65mm,clip=true]{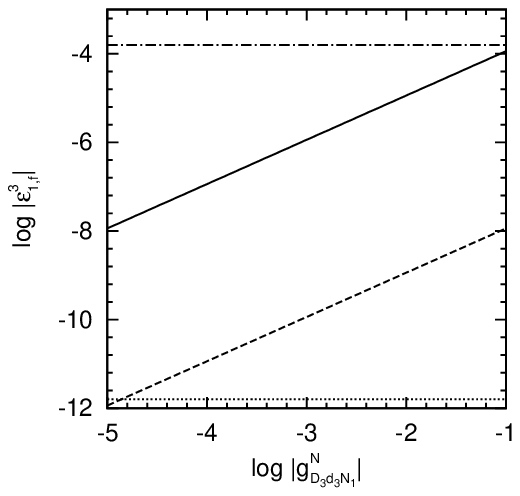}
\hspace*{37mm}{\bf (a)}\hspace*{88mm}{\bf (b) }\\[3mm]
\caption{Maximal absolute values of the CP asymmetries in the E$_6$SSM Model II as a function of {\it (a)} $|g^N_{D_3 d_3 N_2}|$
and {\it (b)} $|g^N_{D_3 d_3 N_1}|$ for $M_1=10^{6}\,\mbox{GeV}$ and $M_2=10\cdot M_1$. All couplings $|h^N_{H^u_k L_4 N_j}|$
and $|h^N_{H^u_{\alpha} L_x N_j}|$ ($\alpha=1,\,2$) are set to zero. The solid, dashed and dotted lines in figure {\it (a)}
represent $|\varepsilon^3_{1,\,\mu}|=|\varepsilon^3_{1,\,\tau}|$ computed for $|g^N_{D_3 d_3 N_1}|=0.1$, $10^{-3}$ and $10^{-5}$
while the dash--dotted line corresponds to $|\varepsilon^3_{1,\,D_3}|$. The solid and dashed lines in figure {\it (b)}
show the dependence of $|\varepsilon^3_{1,\,\mu}|=|\varepsilon^3_{1,\,\tau}|$ on $|g^N_{D_3 d_3 N_1}|$ for
$|g^N_{D_3 d_3 N_2}|=0.1$ and $10^{-5}$ while the dash--dotted and dotted lines correspond to $|\varepsilon^3_{1,\,D_3}|$
calculated for $|g^N_{D_3 d_3 N_2}|=0.1$ and $10^{-5}$ respectively.}
\end{figure}

\newpage
\begin{figure}
\begin{center}
\includegraphics[width=75mm,clip=true]{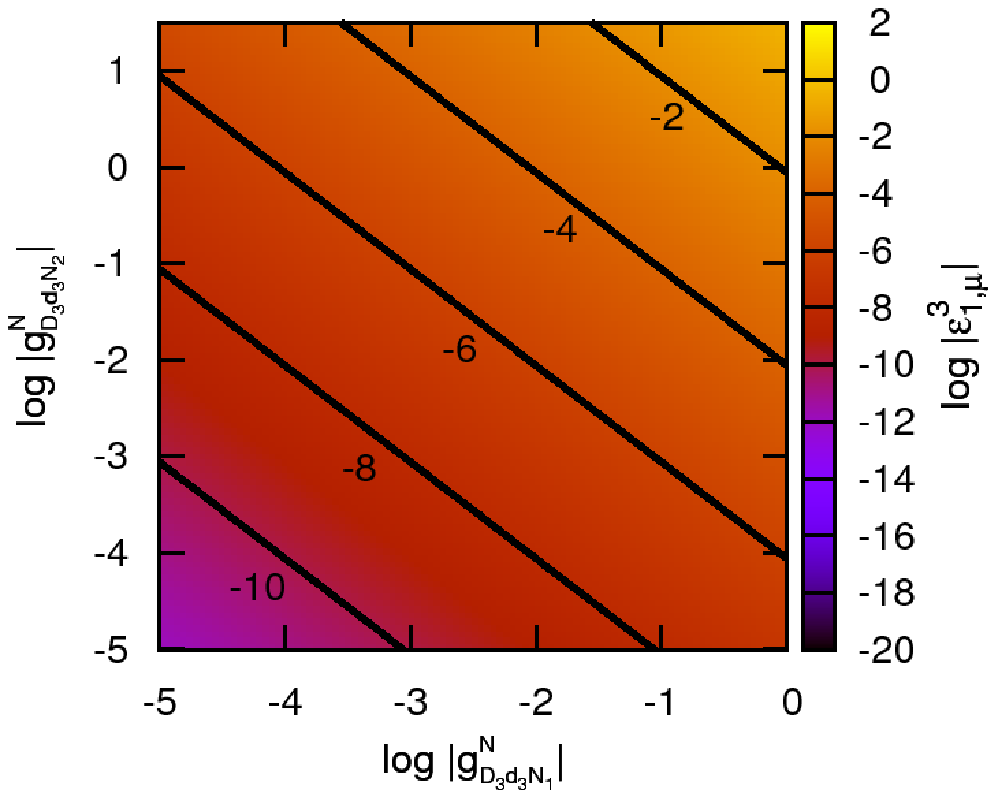} \hspace{5mm}
\includegraphics[width=75mm,clip=true]{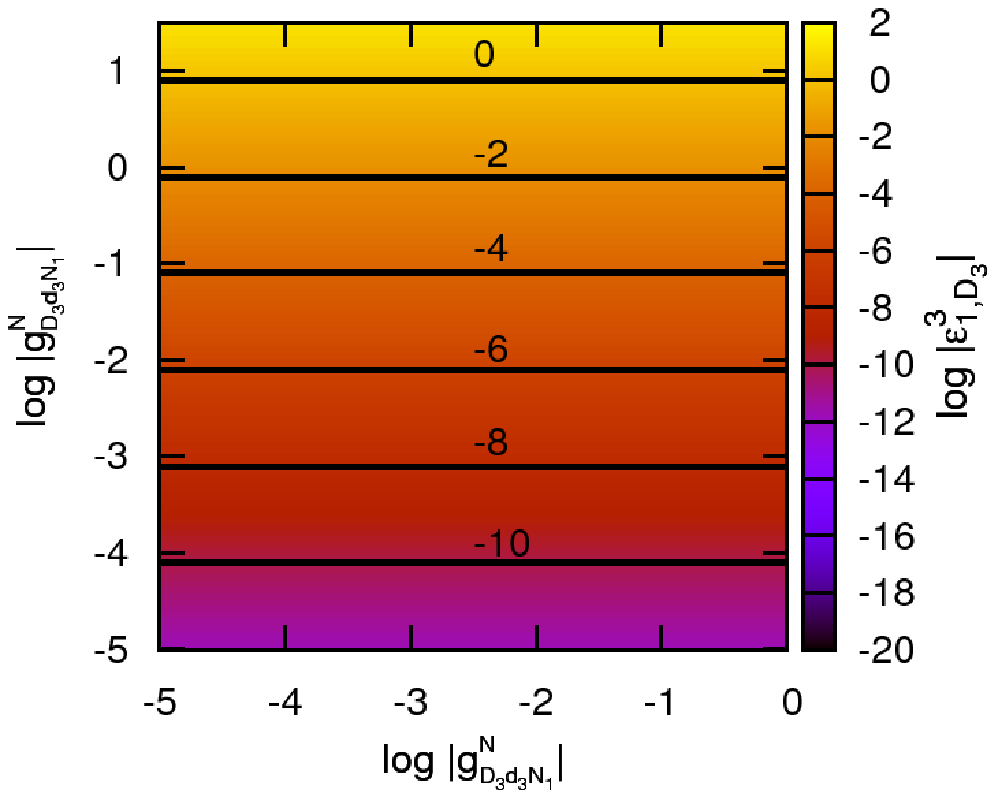}
{\bf (a)}\hspace*{80mm}{\bf (b) }\\[3mm]
\caption{ Logarithm (base 10) of the maximal values of $|\varepsilon^3_{1,\,\mu}|=|\varepsilon^3_{1,\,\tau}|$ ({\it a})
and $|\varepsilon^3_{1,\,D_3}|$ ({\it b}) in the E$_6$SSM Model II versus $\log |g^N_{D_3 d_3 N_2}|$ and $\log |g^N_{D_3 d_3 N_1}|$
for $M_2=10\, M_1$ (fixing the ratio $M_1/M_2$ these asymmetries become independent of $M_1$). All couplings $|h^N_{H^u_k L_4 N_j}|$ and
$|h^N_{H^u_{\alpha} L_x N_j}|$ ($\alpha=1,\,2$) are set to zero. The solid contour lines show steps of $2$ in the logarithm (base 10)
of the asymmetries.}
\end{center}
\end{figure}


\begin{thebibliography}{99}

\bibitem{Komatsu:2008hk}
  E.~Komatsu {\it et al.}  [WMAP Collaboration],
  arXiv:0803.0547 [astro-ph].



\bibitem{gut-baryogen}
A.~Yu.~Ignatiev, N.~V.~Krasnikov, V.~A.~Kuzmin, and A.~N.~Tavkhelidze,
Phys.\ Lett.\ B {\bf 76} (1978) 436;
M.~Yoshimura,
Phys.\ Rev.\ Lett.\ {\bf 41} (1978) 281;
D.~Toussaint, S.~B.~Treiman, F.~Wilczek, and A.~Zee,
Phys.\ Rev.\ D {\bf 19} (1979) 1036;
Steven Weinberg,
Phys.\ Rev.\ Lett.\ {\bf 42} (1979) 850;
M.~Yoshimura,
Phys.\ Lett.\ B {\bf 88} (1979) 294;
S.~M.~Barr, G.~Segre and H.~A.~Weldon,
Phys.\ Rev.\ D {\bf 20} (1979) 2494;
D.~V.~Nanopoulos, S.~Weinberg,
Phys.\ Rev.\ D {\bf 20} (1979) 2484;
A.~Yildiz, P.~Cox,
Phys.\ Rev.\ D {\bf 21} (1980) 906.


\bibitem{ew-baryogen}
A.~Riotto and M.~Trodden,
Ann.\ Rev.\ Nucl.\ Part.\ Sci.\ (1999) {\bf 49} 35;
J.~M.~Cline.
hep-ph/0609145.


\bibitem{Huber:2000mg}
S.~J.~Huber and M.~G.~Schmidt,
Nucl.\ Phys.\  B {\bf 606} (2001) 183.


\bibitem{Fukugita:1986hr}
M.~Fukugita and T.~Yanagida,
Phys.\ Lett.\  B {\bf 174} (1986) 45.

\bibitem{review-leptogen}
{\it for a recent review, see } S.~Davidson, E.~Nardi and Y.~Nir,
arXiv:0802.2962\,.


\bibitem{Affleck-Dine}
I.~Affleck and M.~Dine.
Nucl.\ Phys.\ B {\bf 249} (1985) 361;
M.~Dine, L.~Randall and S.~D.~Thomas,
Nucl.\ Phys.\ B {\bf 458} (1996) 291.


\bibitem{see-saw}
P.~Minkowski, Phys.\ Lett.\ B {\bf 67} (1977) 421;
M.~Gell-Mann, P.~Ramond and R.~Slansky, {\it Proceedings of the Supergravity Stony Brook Workshop},
New York 1979, eds. P.~Van Nieuwenhuizen and D.~Freedman; T.~Yanagida, {\it Proceedings of the Workshop on
Unified Theory and Baryon Number in the Universe}, Tsukuba, Japan 1979, eds. A.~Sawada and A.~Sugamoto;
R.~N.~Mohapatra, G.~Senjanovic, Phys.\ Rev.\ Lett. {\bf 44} (1980) 912.


\bibitem{Sakharov:1967dj}
A.~D.~Sakharov,
Pisma Zh.\ Eksp.\ Teor.\ Fiz.\  {\bf 5} (1967) 32
[JETP Lett.\  {\bf 5} (1967) 24].


\bibitem{Kuzmin:1985mm}
V.~A.~Kuzmin, V.~A.~Rubakov and M.~E.~Shaposhnikov,
Phys.\ Lett.\  B {\bf 155} (1985) 36;
V.~A.~Rubakov and M.~E.~Shaposhnikov,
Usp.\ Fiz.\ Nauk, {\bf 166} (1996) 493.



\bibitem{CPasym-SM}
M.~A.~Luty,
Phys.\ Rev.\  D {\bf 45} (1992) 455;
M.~Flanz, E.~A.~Paschos and U.~Sarkar,
Phys.\ Lett.\  B {\bf 345} (1995) 248
[Erratum-ibid.\  B {\bf 382} (1996) 447];
M.~Plumacher,
Z.\ Phys.\  C {\bf 74} (1997) 549;
W.~Buchmuller and M.~Plumacher,
Phys.\ Lett.\  B {\bf 431} (1998) 354.


\bibitem{CPasym-SUSY}
B.~A.~Campbell, S.~Davidson and K.~A.~Olive,
Nucl.\ Phys.\  B {\bf 399} (1993) 111;
L.~Covi, E.~Roulet and F.~Vissani,
Phys.\ Lett.\  B {\bf 384} (1996) 169;
M.~Plumacher,
Nucl.\ Phys.\  B {\bf 530} (1998) 207.


\bibitem{Buchmuller:2004nz}
G.~F.~Giudice, A.~Notari, M.~Raidal, A.~Riotto, and A.~Strumia,
Nucl.\ Phys.\ B {\bf 685} (2004) 89;
W.~Buchmuller, P.~Di~Bari, and M.~Plumacher,
Ann.\ Phys.\ {\bf 315} (2005) 305.


\bibitem{CPleptogen-flavour}
R.~Barbieri, P.~Creminelli, A.~Strumia, and N.~Tetradis,
Nucl.\ Phys.\ B {\bf 575} (2000) 61;
T.~Endoh, T.~Morozumi, and Z.~h.~Xiong,
Prog.\ Theor.\ Phys.\ {\bf 111} (2004) 123;
O.~Vives,
Phys.\ Rev.\  D {\bf 73} (2006) 073006;
A.~Abada, S.~Davidson, F.~X.~Josse-Michaux, M.~Losada and A.~Riotto,
JCAP {\bf 0604} (2006) 004;
E.~Nardi, Y.~Nir, E.~Roulet and J.~Racker,
JHEP {\bf 0601} (2006) 164;
A.~Abada, S.~Davidson, A.~Ibarra, F.~X.~Josse-Michaux, M.~Losada and A.~Riotto,
JHEP {\bf 0609} (2006) 010;
A.~De Simone and A.~Riotto,
JCAP {\bf 0702} (2007) 005;
S.~Blanchet, P.~Di Bari and G.~G.~Raffelt,
JCAP {\bf 0703} (2007) 012.


\bibitem{Antusch:2006cw}
S.~Antusch, S.~F.~King and A.~Riotto,
JCAP {\bf 0611} (2006) 011.


\bibitem{triplet-seesaw-leptogen1}
E.~Ma and U.~Sarkar,
Phys.\ Rev.\ Lett.\ {\bf 80} (1998) 5716;
T.~Hambye, E.~Ma, and U.~Sarkar,
Nucl.\ Phys.\ B {\bf 602} (2001) 23;
E.~J.~Chun and S.~K.~Kang,
Phys.\ Rev.\ D {\bf 63} (2001) 097902;
A.~S.~Joshipura, E.~A.~Paschos, and W.~Rodejohann,
\newblock {\em Nucl. Phys.}, B611:227--238, 2001;
B.~Brahmachari, E.~Ma, and U.~Sarkar,
Phys.\ Lett.\ B {\bf 520} (2001) 152;
T.~Hambye and G.~Senjanovic,
Phys.\ Lett.\ B {\bf 582} (2004) 73;
W.~l.~Guo,
Phys.\ Rev.\  D {\bf 70} (2004) 053009;
S.~Antusch and S.~F. King,
Phys.\ Lett.\ B {\bf 597} (2004) 199;
S.~Antusch and S.~F.~King,
JHEP\ {\bf 0601} (2006) 117;
T.~Hambye, M.~Raidal, and A.~Strumia,
Phys.\ Lett.\ B {\bf 632} (2006) 667;
E.~J.~Chun and S.~Scopel,
Phys.\ Rev.\  D {\bf 75} (2007) 023508;
S.~Antusch,
Phys.\ Rev.\ D {\bf 76} (2007) 023512;
W.~Chao, S.~Luo and Z.~z.~Xing,
Phys.\ Lett.\  B {\bf 659} (2008) 281;
T.~Hallgren, T.~Konstandin and T.~Ohlsson,
JCAP {\bf 0801} (2008) 014.


\bibitem{triplet-seesaw-leptogen2}
G.~D'Ambrosio, T.~Hambye, A.~Hektor, M.~Raidal, and A.~Rossi,
Phys.\ Lett.\ B {\bf 604} (2004) 199;
E.~J.~Chun and S.~Scopel,
Phys.\ Lett.\  B {\bf 636} (2006) 278.



\bibitem{dirac-leptogen}
K.~Dick, M.~Lindner, M.~Ratz, and D.~Wright,
Phys.\ Rev.\ Lett.\ {\bf 84} (2000) 4039;
H.~Murayama and A.~Pierce,
Phys.\ Rev.\ Lett.\ {\bf 89} (2002) 271601;
M.~Boz and N.~K.~Pak,
Eur.\ Phys.\ J.\ C {\bf 37} (2004) 507;
D.~G.~Cerdeno, A.~Dedes, and T.~E.~J. Underwood,
JHEP\ {\bf 0609} (2006) 067;
B.~Thomas and M.~Toharia,
Phys.\ Rev.\ D {\bf 73} (2006) 063512;
B.~Thomas and M.~Toharia,
Phys.\ Rev.\ D {\bf 75} (2007) 013013.


\bibitem{lower-bound}
S.~Davidson and A.~Ibarra,
Phys.\ Lett.\  B {\bf 535} (2002) 25;
K.~Hamaguchi, H.~Murayama and T.~Yanagida,
Phys.\ Rev.\  D {\bf 65} (2002) 043512.


\bibitem{gravitino-problem}
M.~Yu.~Khlopov and A.~D.~Linde,
Phys.\ Lett.\ B {\bf 138} (1984) 265;
J.~R.~Ellis, J~E.~Kim, and D.~V.~Nanopoulos,
Phys.\ Lett.\ B {\bf 145} (1984) 181.


\bibitem{Kohri:2005wn}
M.~Y.~Khlopov, Yu.~L.~Levitan, E.~V.~Sedelnikov and I.~M.~Sobol,
Phys.\ Atom.\ Nucl.\  {\bf 57} (1994) 1393
[Yad.\ Fiz.\  {\bf 57} (1994) 1466];
M.~Kawasaki, K.~Kohri and T.~Moroi,
Phys.\ Rev.\  D {\bf 71} (2005) 083502;
K.~Kohri, T.~Moroi and A.~Yotsuyanagi,
Phys.\ Rev.\  D {\bf 73} (2006) 123511.


\bibitem{soft-leptogenesis}
Y.~Grossman, T.~Kashti, Y.~Nir and E.~Roulet,
Phys.\ Rev.\ Lett.\  {\bf 91} (2003) 251801;
G.~D'Ambrosio, G.~F.~Giudice and M.~Raidal,
Phys.\ Lett.\  B {\bf 575} (2003) 75;
Y.~Grossman, T.~Kashti, Y.~Nir and E.~Roulet,
JHEP {\bf 0411} (2004) 080;
E.~J.~Chun,
Phys.\ Rev.\ D {\bf 69} (2004) 117303;
L.~Boubekeur, T.~Hambye and G.~Senjanovic,
Phys.\ Rev.\ Lett.\ {\bf 93} (2004) 111601;
M.~C.~Chen and K.~T.~Mahanthappa,
Phys.\ Rev.\  D {\bf 70} (2004) 113013;
T.~Kashti,
Phys. Rev. D {\bf 71} (2005) 013008;
Y.~Grossman, R.~Kitano and H.~Murayama,
JHEP {\bf 0506} (2005) 058;
A.~D.~Medina and C.~E.~M.~Wagner,
JHEP {\bf 0612} (2006) 037;
J.~Garayoa, M.~C.~Gonzalez-Garcia and N.~Rius,
JHEP {\bf 0702} (2007) 021;
E.~J.~Chun and L.~Velasco-Sevilla,
JHEP {\bf 0708} (2007) 075.
S.~Dar, S.~J.~Huber, V.~N.~Senoguz and Q.~Shafi,
Phys.\ Rev.\  D {\bf 69} (2004) 077701;


\bibitem{resonant-leptogen}
A.~Pilaftsis and T.~E.~J.~Underwood,
Nucl.\ Phys.\ B {\bf 692} (2004) 303;
T.~Hambye, J.~March-Russell and S.~M.~West,
JHEP {\bf 0407} (2004) 070;
C.~H.~Albright and S.~M.~Barr,
Phys.\ Rev.\ D {\bf 70} (2004) 033013;
S.~Dar, S.~J.~Huber, V.~N.~Senoguz and Q.~Shafi,
Phys.\ Rev.\  D {\bf 69} (2004) 077701;
A.~Pilaftsis and T.~E.~J.~Underwood,
Phys.\ Rev.\ D {\bf 72} (2005) 113001;
A.~Pilaftsis,
Phys.\ Rev.\ Lett.\ {\bf 95} (2005) 081602;
A.~Anisimov, A.~Broncano and M.~Plumacher,
Nucl.\ Phys.\ B {\bf 737} (2006) 176;
S.~M.~West,
Mod.\ Phys.\ Lett.\ A {\bf 21} (2006) 1629;
Z.~z.~Xing and S.~Zhou,
Phys.\ Lett.\  B {\bf 653} (2007) 278;
A.~De Simone and A.~Riotto,
JCAP {\bf 0708} (2007) 013;
V.~Cirigliano, A.~De Simone, G.~Isidori, I.~Masina and A.~Riotto,
JCAP {\bf 0801} (2008) 004.




\bibitem{new-particles}
T.~Hambye,
Nucl.\ Phys.\  B {\bf 633} (2002) 171;
A.~Abada and M.~Losada,
Nucl.\ Phys.\  B {\bf 673} (2003) 319;
A.~Abada, H.~Aissaoui and M.~Losada,
Nucl.\ Phys.\  B {\bf 728} (2005) 55;
A.~Abada, G.~Bhattacharyya and M.~Losada,
Phys.\ Rev.\  D {\bf 73} (2006) 033006;
D.~Atwood, S.~Bar-Shalom and A.~Soni,
Phys.\ Lett.\  B {\bf 635} (2006) 112;
M.~Frigerio, T.~Hambye and E.~Ma,
JCAP {\bf 0609} (2006) 009;
E.~Ma, N.~Sahu and U.~Sarkar,
J.\ Phys.\ G {\bf 34} (2007) 741;
M.~Hirsch, J.~W.~F.~Valle, M.~Malinsky, J.~C.~Romao and U.~Sarkar,
Phys.\ Rev.\  D {\bf 75} (2007) 011701;
N.~Sahu and U.~Sarkar,
Phys.\ Rev.\  D {\bf 76} (2007) 045014.


\bibitem{inflaton-leptogen}
G.~Lazarides and Q.~Shafi,
Phys.\ Lett.\ B {\bf 258} (1991) 305;
T.~Asaka, K.~Hamaguchi, M.~Kawasaki and T.~Yanagida.
Phys.\ Lett.\ B {\bf 464} (1999) 12.


\bibitem{Giudice:1999fb}
G.~F.~Giudice, M.~Peloso, A.~Riotto and I.~Tkachev,
JHEP {\bf 9908} (1999) 014.


\bibitem{gravitino-LSP}
M.~Bolz, W.~Buchmuller and M.~Plumacher,
Phys.\ Lett.\  B {\bf 443} (1998) 209;
J.~L.~Feng, S.~Su and F.~Takayama,
Phys.\ Rev.\  D {\bf 70} (2004) 075019;
T.~Kanzaki, M.~Kawasaki, K.~Kohri and T.~Moroi,
Phys.\ Rev.\  D {\bf 75} (2007) 025011.


\bibitem{Ibe:2004tg}
M.~Ibe, R.~Kitano, H.~Murayama and T.~Yanagida,
Phys.\ Rev.\ D {\bf 70} (2004) 075012.


\bibitem{King:2005jy}
S.~F.~King, S.~Moretti and R.~Nevzorov,
Phys.\ Rev.\ D {\bf 73} (2006) 035009.


\bibitem{King:2005my}
S.~F.~King, S.~Moretti and R.~Nevzorov,
Phys.\ Lett.\ B {\bf 634} (2006) 278.


\bibitem{Hambye:2000bn}
T.~Hambye, E.~Ma, M.~Raidal and U.~Sarkar,
Phys.\ Lett.\  B {\bf 512} (2001) 373.


\bibitem{SD}
S.~F.~King,
Phys.\ Lett.\  B {\bf 439} (1998) 350;
S.~F.~King,
Nucl.\ Phys.\  B {\bf 562} (1999) 57;
S.~F.~King,
Nucl.\ Phys.\  B {\bf 576} (2000) 85;
S.~F.~King,
JHEP {\bf 0209} (2002) 011.


\bibitem{King:2002nf}
S.~F.~King,
JHEP {\bf 0209} (2002) 011.


\bibitem{King:2002qh}
S.~F.~King,
Phys.\ Rev.\  D {\bf 67} (2003) 113010.


\bibitem{121} Y.~Kawamura,
Prog.\ Theor.\ Phys.\  {\bf 105} (2001) 999;
G.~Altarelli and F.~Feruglio,
Phys.\ Lett.\  B {\bf 511} (2001) 257;
L.~J.~Hall and Y.~Nomura,
Phys.\ Rev.\  D {\bf 64} (2001) 055003;
A.~Hebecker and J.~March-Russell,
Nucl.\ Phys.\  B {\bf 613} (2001) 3;
T.~Asaka, W.~Buchmuller and L.~Covi,
Phys.\ Lett.\  B {\bf 523} (2001) 199;
L.~J.~Hall, Y.~Nomura, T.~Okui and D.~R.~Smith,
Phys.\ Rev.\  D {\bf 65} (2002) 035008.


\bibitem{Accomando:2006ga}
S. Kraml {\it et al.} (eds.), {\it Workshop on CP studies and
non-standard Higgs physics}, CERN--2006--009, hep-ph/0608079;
S.F. King, S. Moretti, R. Nevzorov, hep-ph/0601269;
S.F. King, S. Moretti, R. Nevzorov, hep-ph/0610002;
R.~Howl and S.~F.~King,
JHEP {\bf 0801} (2008) 030.


\bibitem{King:2007uj}
S.~F.~King, S.~Moretti and R.~Nevzorov,
Phys.\ Lett.\  B {\bf 650} (2007) 57.


\bibitem{Howl:2008xz}
  R.~Howl and S.~F.~King,
  arXiv:0802.1909 [hep-ph].


\bibitem{Hemmick:1989ns}
J. Rich, M. Spiro, J. Lloyd-Owen, Phys. Rept. {\bf 151} (1987) 239;
P.F. Smith, Contemp. Phys. {\bf 29} (1988) 159; T.K. Hemmick {\it et al.},
Phys. Rev. D {\bf 41} (1990) 2074.

\bibitem{RadCor}
K.~S.~Babu, C.~N.~Leung and J.~T.~Pantaleone,
Phys.\ Lett.\  B {\bf 319} (1993) 191 [arXiv:hep-ph/9309223];
J.~R.~Ellis, G.~K.~Leontaris, S.~Lola and D.~V.~Nanopoulos,
Eur.\ Phys.\ J.\  C {\bf 9} (1999) 389 [arXiv:hep-ph/9808251];
A.~Dighe, S.~Goswami and P.~Roy,
Phys.\ Rev.\  D {\bf 76} (2007) 096005.

\bibitem{PMNS}
B.~Pontecorvo, Zh. Eksp. Teor. Fiz. {\bf 33} (1957) 549; {\it ibid.} {\bf 34} (1958) 247;
{\it ibid.} {\bf 53} (1967) 1717; Z.~Maki, M.~Nakagawa and S.~Sakata, Prog. Theor. Phys.
{\bf 28} (1962) 870.

\bibitem{Apollonio:1999ae}
M.~Apollonio {\it et al.}  [CHOOZ Collaboration],
Phys.\ Lett.\  B {\bf 466} (1999) 415.


\bibitem{King:2000hk}
S.~F.~King and N.~N.~Singh,
Nucl.\ Phys.\  B {\bf 591} (2000) 3.

\bibitem{tribimax}
P.~F.~Harrison, D.~H.~Perkins and W.~G.~Scott,
Phys.\ Lett.\  B {\bf 458} (1999) 79;
P.~F.~Harrison, D.~H.~Perkins and W.~G.~Scott,
Phys.\ Lett.\  B {\bf 530} (2002) 167;
P.~F.~Harrison and W.~G.~Scott,
Phys.\ Lett.\  B {\bf 535} (2002) 163;
P.~F.~Harrison and W.~G.~Scott,
Phys.\ Lett.\  B {\bf 557} (2003) 76;
C.~I.~Low and R.~R.~Volkas,
Phys.\ Rev.\  D {\bf 68} (2003) 033007.
A similar but physically different form was proposed earlier by:
L.~Wolfenstein, Phys.\ Rev.\ D {\bf 18} (1978) 958.

\bibitem{King:2005bj}
S.~F.~King,
JHEP {\bf 0508} (2005) 105.


\bibitem{Fogli:2005cq}
G.~L.~Fogli, E.~Lisi, A.~Marrone and A.~Palazzo,
Prog.\ Part.\ Nucl.\ Phys.\  {\bf 57} (2006) 742.





\end{thebibliography}
\end{document}